\documentclass[12pt]{article}
\usepackage{graphicx, amsmath, amssymb, cite, setspace, color, relsize, bm, bbold, array}

\usepackage[top=1 in, bottom=1 in, left=.9 in, right=.9 in]{geometry}

\newcommand{\captionfonts}{\footnotesize} 
\makeatletter
\long\def\@makecaption#1#2{%
  \vskip\abovecaptionskip
  \sbox\@tempboxa{{\captionfonts #1: #2}}%
  \ifdim \wd\@tempboxa >\hsize
    {\captionfonts #1: #2\par}
  \else
    \hbox to\hsize{\hfil\box\@tempboxa\hfil}%
  \fi
  \vskip\belowcaptionskip}
\makeatother

\def\lsim{ \lower .75ex \hbox{$\sim$} \llap{\raise .27ex
\hbox{$<$}} }
\def\gsim{ \lower .75ex \hbox{$\sim$} \llap{\raise .27ex
\hbox{$>$}} }

\newcommand\T{\rule{0pt}{4.5ex}}       
\newcommand\B{\rule[-3.0ex]{0pt}{0pt}} 

\def\fnote#1#2{\begingroup\def\thefootnote{#1}\footnote{#2}
     \addtocounter{footnote}{-1}\endgroup}

\usepackage{titlesec}
\titleformat{\paragraph}
{\normalfont\normalsize\bfseries}{\theparagraph}{1em}{}
\titlespacing*{\paragraph}
{0pt}{3.25ex plus 1ex minus .2ex}{1.5ex plus .2ex}

\let\oldsqrt\sqrt
\def\sqrt{\mathpalette\DHLhksqrt}
\def\DHLhksqrt#1#2{%
\setbox0=\hbox{$#1\oldsqrt{#2\,}$}\dimen0=\ht0
\advance\dimen0-0.2\ht0
\setbox2=\hbox{\vrule height\ht0 depth -\dimen0}%
{\box0\lower0.4pt\box2}}

\begin{document}

\title{Flux Compactifications on $(S_2)^N$}

\author{Adam~R.~Brown$^{1}$, Alex~Dahlen$^{2}$ and Ali Masoumi$^{3}$ \vspace{.1 in}\\
 \vspace{-.3 em}  $^1$ \textit{\small{Physics Department, Stanford University, Stanford, CA 94305, USA}} \\
 \vspace{-.3 em}  $^2$ \textit{\small{Berkeley Center for Theoretical Physics, Berkeley, CA 94720, USA}}  \\
  \vspace{-.3 em}  $^3$ \textit{\small{Tufts Institute of Cosmology, Medford, MA 02155, USA
}} }
\date{}

\maketitle
\fnote{}{\hspace{-.65cm}emails: \tt{adambro@stanford.edu, adahlen@berkeley.edu, ali@cosmos.phy.tufts.edu}}
\vspace{-.95cm}

\maketitle
\begin{abstract}

\noindent {We investigate a simple extra-dimensional model and its four-dimensional vacua. This model has a two-form flux and a positive cosmological constant, and the extra dimensions are compactified as the product of $N$ two-spheres. The theory is an interesting laboratory because it is at once simple enough to be soluble but rich enough to exhibit many features not present in previous model landscapes.}

\end{abstract}

\newpage

\section{Introduction}

 Our understanding of the physics of large landscapes is advancing from two directions. 
`Top-down' progress is being made towards understanding the compactified vacua of string theory. This involves extrapolating from that small corner  of the string-theory landscape where computation is tractable \cite{Douglas:2006es,Denef:2004ze,Douglas:2003um,Giryavets:2004zr,Gmeiner:2005vz,Douglas:2006xy,Kumar:2006tn,Ashok:2003gk}. 
`Bottom-up' progress is being made by studying simple landscapes that can be understood completely, and increasing the complexity step by step. This paper takes the latter approach.

Two `bottom-up' models that have received a lot of attention are the Bousso-Polchinski model, which has many cycles for the flux to wrap but non-dynamical extra dimensions, 
and what we will call the $N=1$ Freund-Rubin model, which has only a single cycle, but for which the extra dimensions are dynamical. In this paper we propose a new model  both with many cycles and with dynamical extra dimensions. 
We will see that the combination 
of these two attributes gives rise to an elaborate phase diagram and a distribution of vacua with a special feature at $V_4=0$.

 \begin{center}
\begin{tabular}{|c!{\vrule width 2pt}c|c|c!{\vrule width 2pt}c|} \hline 
   & \begin{tabular}{@{}c@{}} $N=1$ \\ $\Lambda>0$ \end{tabular}    &     \begin{tabular}{@{}c@{}} $N=1$ \\ $\Lambda\le 0$ \end{tabular}     &   \begin{tabular}{@{}c@{}} Bousso- \\ Polchinski \end{tabular}  &   \begin{tabular}{@{}c@{}} $N\ge2$ \\ $\Lambda>0$ \end{tabular}    \T\B  \\ 
\noalign{\hrule height 2pt}
extra dims.  &  &  & & \\
are & \checkmark & \checkmark & \hspace{29mm}  & \checkmark \\
dynamical &  &  & & \\
\hline
dS$_4$ vacua exist & \checkmark &  & \checkmark&\checkmark \T\B \\
  \hline
multiple cycles & \hspace{29mm}    &\hspace{29mm}    & \checkmark & \checkmark \T\B  \\
 \hline
finely  &  &  & & \\
spaced  &  & \checkmark  & \checkmark & \checkmark \\
vacua &  & {\smaller only near $V_4 = 0^{-}$} & & \\
\hline
no limit on  & {\smaller too much flux} &  & {\smaller too much flux } & \\
number of &  {\smaller  leads to }& \checkmark &  {\smaller makes $V_4$} & \checkmark \\
flux units   & {\smaller decompactification}& & {\smaller super-Planckian} &\hspace{29mm}   \\
\hline
accumulation &   &&   &{\smaller dS pile-up at 0$^{+}$} \\
of vacua    &  & \checkmark & & \checkmark \\
at $V_4 = 0$ & & {\smaller AdS pile-up at 0$^{-}$}   & & {\smaller AdS pile-up at 0$^{-}$}   \\
\hline
\end{tabular}
\end{center}
\vspace{-.19in}
\makebox[5.35in][r]{}\vspace{.11in}
\vspace{-.25in} 
\begin{figure}[htbp] 
\caption{Unlike previous model landscapes, the $N\ge2$ Freund-Rubin model has both de Sitter and anti-de Sitter vacua with arbitrarily many flux units wrapped around any given cycle. This gives rise to a  pile-up in the number of vacua both as $V_4 \rightarrow 0^{-}$ and as $V_4 \rightarrow 0^+$.}
  \label{fig:OURSandTHEIRS}
\end{figure}

\begin{figure}[t] 
   \centering
   \includegraphics[width=\textwidth]{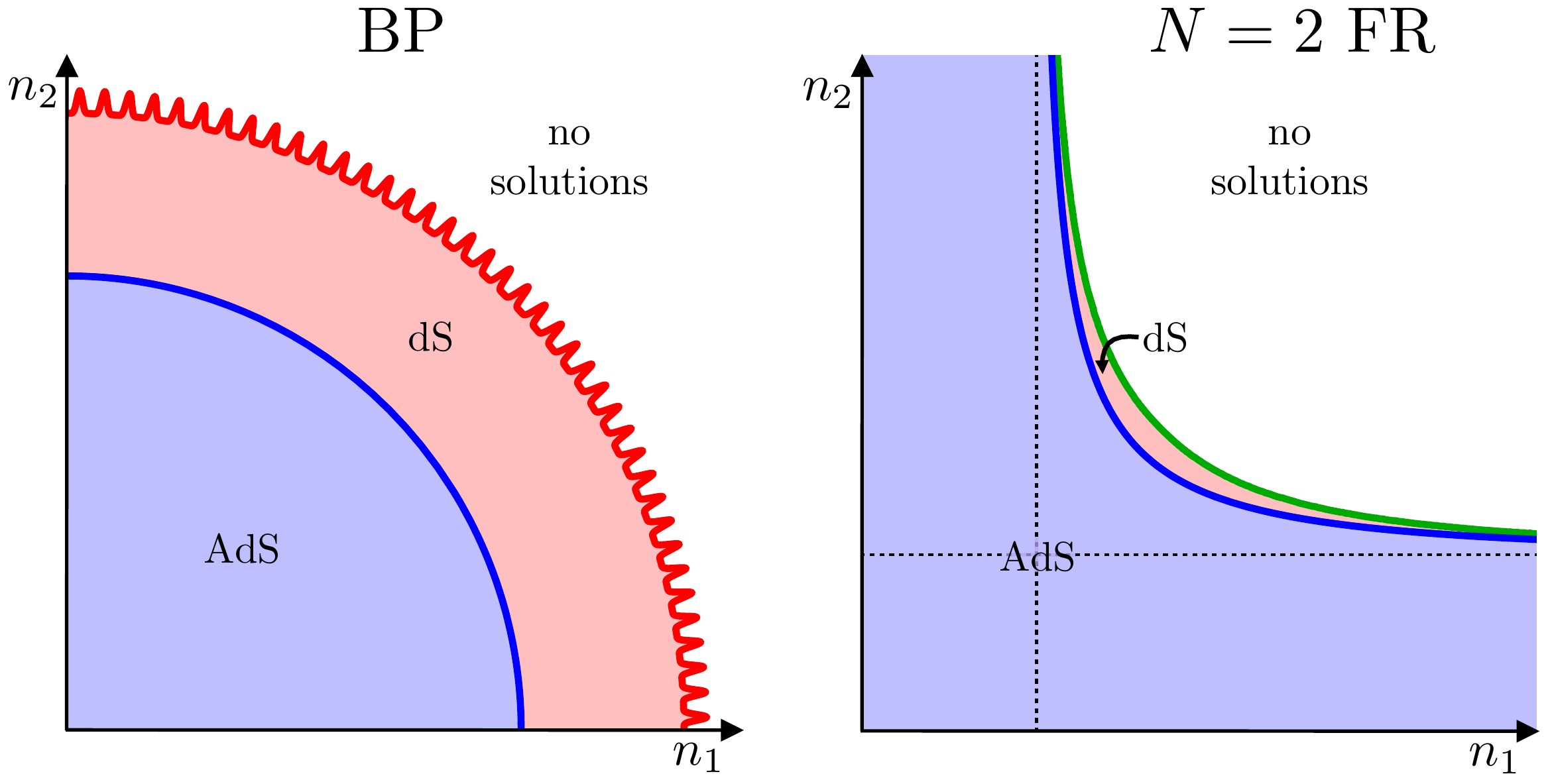}
   \caption{The number of units of flux wrapping each  cycle determines the properties of the corresponding vacuum.  This diagram shows, for the case of $N=2$ cycles, whether the minimum exists and, if so, whether it is AdS, Minkowski, or de Sitter. The negative-$n_i$ quadrants are suppressed in this diagram, and vacua live at Cartesian grid points. On the left is the Bousso-Polchinski model. The line that corresponds to Minkowski minima, given in blue, is a circle. Vacua that lie inside this circle have $V_4 <0$ and are therefore AdS; vacua that lie outside this circle have $V_4 > 0$ and are therefore de Sitter, except that if the flux gets too large (the jagged line) the energy density becomes super-Planckian. On the right is the $N=2$ FR model with higher-dimensional $\Lambda > 0$.  The  Minkowski line, given in blue, is now non-compact,  and asymptotes to the dotted lines. Vacua that lie below the Minkowski curve are AdS.   Once again too much flux can mean no vacuum, but for a reason different than in the BP case---rather than leading to super-Planckian energy densities, too much flux can now lead to decompactification.  The catastrophe line above which there are no longer vacua is shown in green. The crescent between the Minkowski line and the green catastrophe line contains the de Sitter vacua. For this $N=2$ case the crescent has a finite area, but for $N \geq 3$ the volume of $n_i$-space corresponding to de Sitter vacua is infinite; the divergence comes from vacua in which all but one of the cycles have many flux units.   We'll discuss this figure in more detail in Sec.~\ref{N2}.}  
    \label{fig:BPEqui}
\end{figure}

\begin{figure}[t] 
   \centering
   \includegraphics[width=\textwidth]{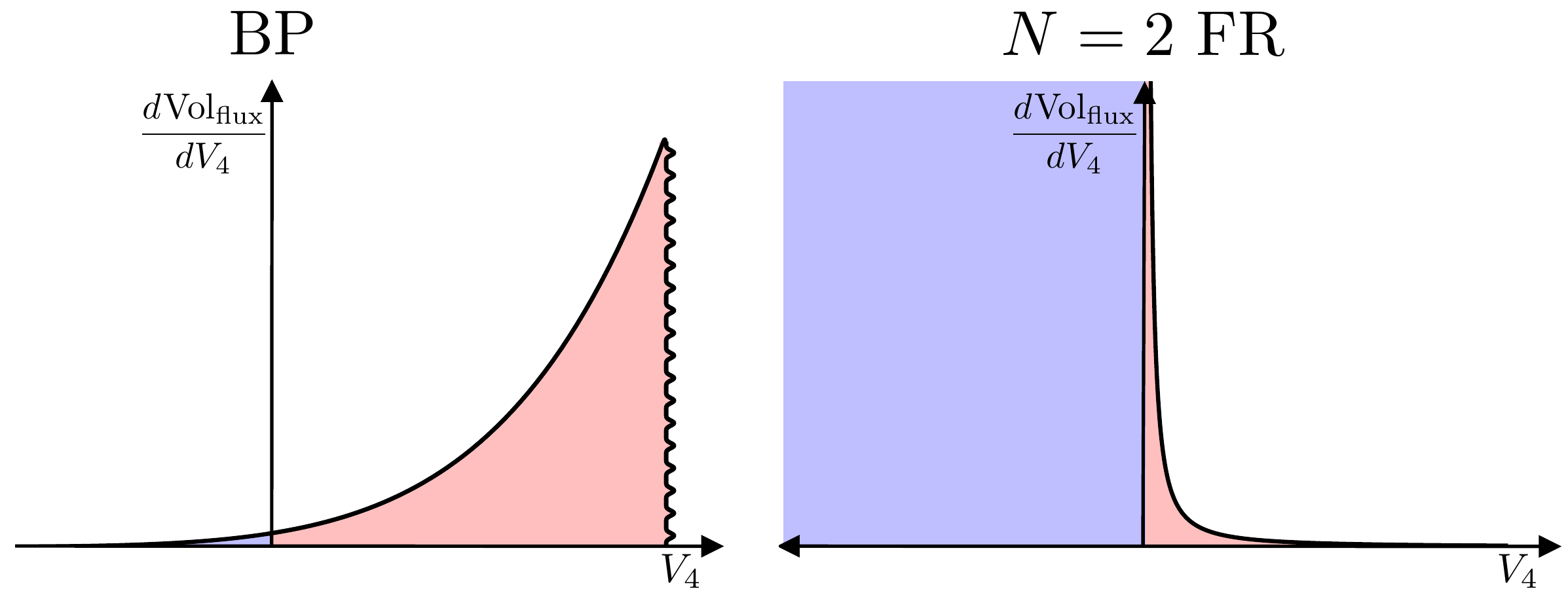}
   \caption{The flux-space volume of vacua with a given 4-dimensional cosmological constant $V_4$, which can be taken as a proxy for the number of such vacua.  The left panel shows the histogram for the BP model.  The distribution is flat through $V_4=0$ and cuts off sharply where the energy density in flux becomes too large. The preponderance of vacua lie just beneath the cutoff. The right panel shows the histogram for the $N\ge2$ FR model; in order to get four-dimensional de Sitter vacua we have taken the higher-dimensional $\Lambda >0$.  There is an unbounded flux-space volume of vacua with any negative value of $V_4$, as well as an accumulation point of de Sitter vacua.}
    \label{fig:BPHisto}
\end{figure}

\subsection{Previous models}
\label{sec:previousmodels}

\textbf{The Bousso-Polchinksi model:}  The Bousso-Polchinski (BP) model \cite{Bousso:2000xa,SchwartzPerlov:2006hi,Clifton:2007bn,Bousso:2007er,Brown:2010mg} is the simplest model with a vast number of vacua; there are many cycles but the extra dimensions are fixed.   

Each vacuum in the theory is labeled by  $N$ integers $(n_1,\dots,n_N)$ that give the number of units of flux around each of the $N$ internal cycles.  The size and shape of the extra dimensions are held fixed by an unspecified (supersymmetry-breaking) sector of the theory, and backreaction of the flux on the extra dimensions is taken to be negligible compared to the contributions of this other sector.  This model has a negative higher-dimensional cosmological constant, which is shifted up by the wrapping fluxes.   The effective four-dimensional cosmological constant $V_4$ in a given vacuum is schematically given by:
\begin{gather}
V_4\sim V_{\textrm{no flux}} + \sum_{i=1}^N g^2 n_i^{\,2},
\end{gather}
where $V_{\textrm{no flux}} < 0$ and $g$ is the unit of magnetic charge. Equipotential surfaces in this theory are $(N-1)$-spheres in $n_i$-space, as shown in Fig.~\ref{fig:BPEqui}.   

Consider the distribution of vacua as a function of $V_4$.  Since the allowed values of the flux are quantized, the  vacua lie at the integer grid points in $n_i$-space.  We can treat the volume in $n_i$-space as a proxy for the number of vacua; the number of grid points in a given region (provided it contains many unit cells) is approximated by the volume of that region divided by the volume of a unit cell. 
 Within the context of this approximation, the number density of vacua $d\hspace{.2mm}\text{Num}/dV_4$ is given by the surface area of the  equipotential sphere with radius $n\sim( V_4-V_{\textrm{no flux}})^{1/2}/g$:
\begin{gather}
\label{BPHisto}
\frac{d\hspace{.2mm}\text{Num}}{dV_4} \sim \frac{( V_4-V_{\textrm{no flux}})^{\frac{N-1}{2}}}{g^{N}}.
\end{gather}
This number density of vacua is exponentially large in $N$ and  plotted in the left panel
  of Fig.~\ref{fig:BPHisto}. Since the equipotential spheres slice through the grid of integer $n_i$'s (rather than aligning with it), the flux-space-volume is an excellent proxy for the number density of minima on all but the finest scales. We expect there to be a minimum exponentially close to any given value of $V_4$.

Because the extra dimensions are held fixed, increasing the number of flux units  increases the higher-dimensional energy density.  When $n$ becomes too large the energy density becomes super-Planckian: effective field theory breaks down for vacua above this cutoff.   Because so much of an $N$-sphere's volume lies at large radius, most vacua in the BP landscape have $V_4$ just below this cutoff.  For example, when $N\ge69$, more than half of the volume lies within $1\%$ of the cutoff.

At all except the finest scales, the distribution of vacua in the BP landscape is smooth through $V_4=0$.  Indeed $V_4=0$ {cannot} be a special location in the BP model, because at fixed size and shape of the extra dimensions, it's non-gravitational physics that sets the effective cosmological constant.

\newpage

\noindent \textbf{The $\bf{N=1}$ Freund-Rubin model:} The $N=1$ Freund-Rubin (FR) model  \cite{Freund:1980xh,RandjbarDaemi:1982hi,BlancoPillado:2009di,Yang:2009wz,Brown:2010bc,Brown:2010mf,BlancoPillado:2009mi} is the simplest self-contained model of flux compactification; the extra dimensions are dynamical, but there's only one cycle. 
A two-form flux is wrapped around an extra-dimensional two-sphere; the sphere's tendency to collapse under its own curvature is counteracted by the repulsive force of the flux. This model is reviewed in Sec.~\ref{N1}.

\subsection{The $\bf{N\ge2}$ Freund-Rubin model}

In this paper, we study the generalization of the Freund-Rubin model to more internal cycles. As in the $N=1$ case, the higher-dimensional theory is endowed with a positive  cosmological constant and a two-form flux. The $2N$ extra dimensions are compactified as the product of $N$ two-spheres, and each sphere is individually wrapped by the  flux. Each vacuum is therefore indexed by $N$ integers that label the number of flux units around each cycle.  As in the BP model, there are many cycles; as in the $N=1$ FR model, the extra dimensions are dynamical.

The $N\ge2$ FR model thus encompasses both previous models. Indeed, we will see that there is a regime of this theory in which it resembles the BP model, another regime in which it closely resembles the $N=1$ FR model with $\Lambda >0$, and yet another regime in which (despite itself having $\Lambda>0$) it closely resembles the $N=1$ FR model with $\Lambda \leq 0$. As well as encompassing previous models, the $N\ge2$ FR model exhibits new phenomena not present in any of them. In particular we will see that there is no bound on the number of units of flux around any one cycle (and indeed the typical de Sitter vacuum has many units of flux around all but one cycle), and we will see that there is a double-exponentially large number of de Sitter vacua whose cosmological constants accumulate at $V_4=0^+$.  \\

\vspace{-.75cm}
\begin{equation}\nonumber
\star \, \, \star \, \,\star
\end{equation}

Accumulation points in the distribution of  vacua have been found before, but they have always previously been for AdS vacua. A simple example  of AdS vacua piling up at  $V_4 = 0^-$ is the $N=1$  FR model with $\Lambda\leq0$, which we will review in Sec.~\ref{N1}. We believe the $N \geq 2$ FR model is the first example of  de Sitter vacua piling-up at $V_4 = 0^+$. (Ashok and Douglas \cite{Ashok:2003gk} have argued that pile-ups of vacua are always associated with decoupling limits in which the internal volume gets very large, and ours will be no exception; this large internal volume dilutes away all contributions to the effective potential \cite{Dine:1985he, Giddings:2003zw, Silverstein:2004id,Braun:2011av,deAlwis:2013gka}.) 

A (1+1)-dimensional version of this landscape was studied in \cite{Asensio:2012pg,Asensio:2012wt}. In \cite{Brown:2013fba}, we analyzed  de Sitter accumulation points more broadly, and argued that they are generic in higher-dimensional theories with de Sitter minima---de Sitter vacua beget more de Sitter vacua.
The full perturbative spectrum of these compactifications was studied in  \cite{Brown:2013mwa}; the important result for the purpose of this paper is that  all higher spherical-harmonic fluctuations are stable and can be safely truncated. (For the special case of $N=1$ the stability had already been established in \cite{DeWolfe:2001nz, Bousso:2002fi} and then again in \cite{Hinterbichler:2013kwa}.) 

In Secs.~\ref{effpotderive}~\&~\ref{extremaequations}, we derive the effective potential for the radii of the $N$ internal spheres and give formulae  for the properties of the vacua, including the size of the extra dimensions, the effective cosmological constant, and the radion mass matrix.  In Sec.~\ref{N1}, we review the $N=1$ Freund-Rubin model as a special case, and  highlight the different large-flux behaviors for negative, zero, and positive $\Lambda$. In Sec.~\ref{N2}, we explore the $N=2$ Freund-Rubin model; this is the simplest case that exhibits our new phenomena, and we map out the full phase diagram. In Sec.~\ref{sec:GeneralN}, we extend this analysis to general $N$. We show that there are a number of regimes in which simple analytic formulae for the properties of the minima can be derived.  In Sec.~\ref{sec:statistics} we discuss the statistics of these vacua and show that  most of the vacua have all but one of the flux numbers huge.

\section{Dimensional Reduction}
\label{effpotderive}
The higher-dimensional action is
\begin{gather}
\label{Daction}
S=\int d^4x d^2y_1\cdots d^2 y_N \sqrt{-G}\left(\frac12M_D^{\,D-2} \,\mathcal{R}_D-\frac{1}{4} M_D^{\,D-4}F^2 - \Lambda_D M_D^{\,D}\right),
\end{gather}
where $G_{MN}$ is the $(D=4+2N)$-dimensional metric, $F_{MN}$ is a two-form flux, $\Lambda_D>0$ is a positive cosmological constant, and $M_D$ is the $D$-dimensional Planck mass. The only dimensionful quantities are $M_D$, which has mass-dimension one, and $G_{MN}$ and $\mathcal{R}_D$, which have mass dimension  minus two; the quantities $F_{MN}$ and $\Lambda_D$ are defined to be dimensionless. The Hubble scale of the $D$-dimensional de Sitter space is $H_D\equiv\sqrt{\Lambda_D}M_D$. 

We are interested in compactified solutions of the form $M_4\times (S_2)^N$; the metric is taken to be a  product
\begin{gather}
\label{metricansatz}
ds^2=\frac{1}{H_D^{\,2}}\left\{\left(\prod_{i=1}^N\frac{1}{R_i(x)^2}\right)g_{\mu\nu}dx^\mu dx^\nu + \sum_{i=1}^N R_i(x)^2 d\Omega_{2,i}^{\,2}\right\},
\end{gather}
where $g_{\mu\nu}$ is a maximally symmetric 4-dimensional metric---either de Sitter, Minkowski, or AdS---and $R_i(x)$ is a 4-dimensional field proportional to the radius of the $i$th 2-sphere.  All distances are measured in units of the higher-dimensional Hubble length $H_D^{-1}$, so $g_{\mu\nu}$ and $R_i$ are both dimensionless.  We've pulled off a Weyl factor (the product of the $R_i^{-2}$s) from the 4D metric; we will see that this step puts the dimensionally reduced theory in Einstein frame. The two-form flux is taken to wrap each 2-sphere uniformly
\begin{gather}
\label{fluxconfiguration}
{\mathbf F} = \left(\sqrt{\frac{2}{\Lambda_D}}\,\right) \sum_{i=1}^N n_i \; \sin \theta_i d\theta_i\wedge d\phi_i,
\end{gather}
where $n_i$ is proportional to the number of flux units wrapping the $i$th 2-sphere, and $\theta_i$ and $\phi_i$ are coordinates on that sphere.  The factor of $\sqrt{2/\Lambda_D}$ in the definition of $n_i$ is there for later convenience. 
The $n_i$ obey a Dirac quantization condition that we will discuss in Sec.~\ref{Dirac}.

After plugging the metric ansatz Eq.~\ref{metricansatz} and the background flux configuration Eq.~\ref{fluxconfiguration} into the $D$-dimensional action  Eq.~\ref{Daction}, and integrating over the $N$ 2-spheres, the action  becomes
\begin{gather}
S=\frac{(4\pi)^N}{\Lambda_D^{\,N+1}}\int d^4 x \sqrt{-g} \Bigg[\frac12\mathcal{R}_4 - \nabla_\mu \left(\sum_{i=1}^N \ln R_i(x)\right)\nabla^\mu\left(\sum_{i=1}^N\ln R_i(x)\right)\nonumber\\ -  \sum_{i=1}^N\left(\nabla_\mu\left[ \ln R_i(x)\right] \nabla^\mu \left[ \ln R_i(x)\right]\right) - V_{}(R_i)\Bigg]; \label{effSR}
\end{gather}
the kinetic term for the $R_i$ fields is non-canonical and couples the fields together.  The effective potential for the $R_i$ fields is
\begin{gather}
\label{effpotR}
\boxed{V_{}(R_i)=\left\{\prod_{i=1}^N\frac{1}{R_i^{\,2}}\right\}\left[\sum_{i=1}^N \frac{n_i^{\,2}}{R_i^{\,4}}-\sum_{i=1}^N\frac1{R_i^{\,2}} + 1\right].}
\end{gather}
Every term in this action is dimensionless, and the only higher-dimensional quantity $\Lambda_D$ appears outside the action as a multiplicative factor.  

The Weyl factor  extracted from the four-dimensional metric $g_{\mu\nu}$ in Eq.~\ref{metricansatz} has put this action in Einstein frame, meaning that there are no factors of the four-dimensional $R_i$ fields that multiply the Einstein term $\mathcal{R}_4$. The Weyl factor can be thought of as the unit conversion factor from the higher-dimensional Planck mass to the 4-dimensional Planck mass, so that the value of the effective potential at a minimum, $V_4$, is  the cosmological constant in 4d Planck units.

We will sometimes write this effective potential as
\begin{gather}
\label{Vtildedef}
V_{}(R_i)=\left\{\prod_{i=1}^N\frac{1}{R_i^{\,2}}\right\} \widetilde V(R_i), \hspace{.3in} \text{where} \hspace{.3in}  \widetilde V(R_i) \equiv \sum_{i=1}^N \frac{n_i^{\,2}}{R_i^{\,4}}-\sum_{i=1}^N\frac1{R_i^{\,2}} + 1.
\end{gather}
The first term in $\widetilde V$ represents the flux contribution: flux lines repel and give a hard-core repulsive force that buttresses the spheres against collapse.  The second term in $\widetilde V$ represents the spatial-curvature contribution of the spheres, which makes them want to shrink to zero size.  The last term in $\widetilde V$ represents the cosmological-constant contribution, as can be made manifest by `un-scaling' $R_i$ and $n_i$ by $\sqrt{\Lambda_D}$, and $\widetilde V$ by $\Lambda_D^{-1}$.

\subsection{Canonically normalized kinetic terms}
\label{CanonKinetic}
The kinetic term for the $R_i$ in the effective action Eq.~\ref{effSR} is non-canonical. If we define
\begin{gather}
A=\frac{\sqrt{N+1}-1}{N\sqrt{N+1}},
\label{eq:Adefintion}
\end{gather}
then canonically normalized coordinates are given by
\begin{gather}
\ln R_i(x)=\frac1{\sqrt2} \phi_i(x) - \frac1{\sqrt2} A \sum_{k=1}^N\phi_k(x).
\end{gather}
The inverse relation is
\begin{gather}
\label{phiofR}
\phi_i(x) = \sqrt2\ln R_i(x) + \sqrt2\, \sqrt{N+1} \,A \sum_{k=1}^N \ln R_k(x).
\end{gather}
These $\phi_i$ correspond to combinations of the $R_i$ in which one of the spheres increases in radius while all the others shrink. In terms of these canonical coordinates, the metric ansatz of Eq.~\ref{metricansatz} becomes
\begin{gather}
ds^2=\frac{1}{H_D^{\;2}}\left\{\left(e^{\sqrt2(NA-1)\sum_{k=1}^N \phi_k(x)}\right)g^{\mu\nu}dx_\mu dx_\nu + \left(e^{-\sqrt2A \sum_{k=1}^N\phi_k(x)} \right)\sum_{i=1}^N e^{\sqrt2\phi_i(x)} d\Omega_{2,i}^{\,2}\right\},
\end{gather}
and the action of Eq.~\ref{effSR} becomes
\begin{gather}
S_{{}}=\frac{(4\pi)^N}{\Lambda_D^{\,N+1}}\int d^4 x \sqrt{-g} \Bigg[\frac12\mathcal{R}_4 -\frac12 \sum_{i=1}^N \nabla_\mu \phi_i(x)\nabla^\mu\phi_i(x)- V_{{}}(\phi_i)\Bigg].
\end{gather}
As promised, the kinetic term for the $\phi_i$ fields is canonical.  The effective potential in terms of these new coordinates becomes:
\begin{gather}
V_{}(\phi_i)=e^{\sqrt2(NA-1)\sum_{j=1}^N \phi_j}\left[e^{2\sqrt2 A\sum_{k=1}^N\phi_k}\sum_{i=1}^N n_i^{\,2}e^{-2\sqrt2\phi_i}-e^{\sqrt2A\sum_{k=1}^N\phi_k}\sum_{i=1}^N  e^{-\sqrt2\phi_i} + 1\right].
\end{gather}

To calculate the radion masses, we will need to use canonical coordinates and differentiate $V(\phi_i)$. However, since the coordinate transformation that relates the $\phi_i$ to the $R_i$ is positive-definite, the critical points of $V(R_i)$ are the same as the critical points of  $V(\phi_i)$, and have the same number of negative eigenmodes.

\subsection{The sign of $\mathbf{\Lambda_D}$}
This paper is concerned with the $N\ge2$ FR model with $\Lambda_D$  positive. However, the analysis above can be extended to the case where $\Lambda_D$ is negative. Since we scaled out the definition of $R_i$ and $n_i$ by $\sqrt{\Lambda_D}$, this involves analytically continuing $R_j\rightarrow i R_j$ and $n_j\rightarrow i n_j$, which preserves the positivity of the metric and background flux ansatze in Eqs.~\ref{metricansatz} and \ref{fluxconfiguration}.  
The effective potential for either sign of $\Lambda_D$ can then be written as
\begin{gather}
\label{effpotRLambdaneg}
V_{}(R_i)= \left\{\prod_{i=1}^N\frac{1}{R_i^{\,2}}\right\} \left[\sum_{i=1}^N \frac{n_i^{\,2}}{R_i^{\,4}}-\sum_{i=1}^N\frac{1}{R_i^{\,2}} + \text{sign}(\Lambda_D)\right].
\end{gather}
When $\Lambda_D = 0$, sign($\Lambda_D) = 0$ and the dependencies on $\Lambda_D$ of $n_i$, $R_i$ and the four-dimensional Planck mass cancel in the effective potential. 
The $\Lambda_D \leq 0$ case is of less interest because there is only ever a single extremum and it is always an AdS minimum: when $\Lambda_D \leq 0$ the only positive contribution to the effective potential is the flux term.

\section{Extrema, Catastrophes, and Radion Masses}
 \label{extremaequations}

In this section we look at critical points of the effective potential, Eq.~\ref{Vtildedef}, and the mass matrix of small fluctuations around them. We pay particular attention to the minima, which give the vacua of the theory.  

\subsection{Extrema}
\label{sec:extrema}
To find extrema of the effective potential, it is easiest to work in terms of the non-canonical coordinates $R_i$, for which the effective potential is given in Eq.~\ref{Vtildedef}.  For all $i$, extrema satisfy
\begin{gather}
\frac{\partial V_{}}{\partial R_i}=V_{}\left(-\frac{2}{R_i} + \frac{\partial \widetilde V}{\partial R_i} \frac1{\widetilde V}\right)=0 \ \ \leftrightarrow \ \
\frac{R_i^{\,2}-2n_i^{\,2}}{R_i^{\,4}} = \widetilde V  \ \ \leftrightarrow \  \ R_i^{\,2}=\frac{1\pm\sqrt{1-8n_i^{\,2}\widetilde V}}{2\widetilde V} . \label{extremum}
\end{gather}
Minkowski extrema, in addition to satisfying Eq.~\ref{extremum}, also have $V=\widetilde V=0$, so that
\begin{gather}
\label{MinkN}
\text{Minkowski:}\hspace{.65in}  R_i^{\,2}=2 n_i^{\,2}, \hspace{.5in} \sum_{i=1}^N\frac1{n_i^{\,2}}=4.
\end{gather}
Other extrema satisfy:
\begin{gather}
\label{dSN}
\text{de Sitter:}\hspace{.2in}  R_i^{\,2}>2 n_i^{\,2}, \hspace{.8in}\text{and}\hspace{.8in} \text{AdS:}\hspace{.2in}R_i^{\,2}<2 n_i^{\,2},
\end{gather}
for all $i$.
When $N=1$, the Minkowski minimum equation, Eq.~\ref{MinkN},  has only one solution $|n_1|=1/2$. But when $N\ge2$, there is an entire codimension-one surface of $n_i$'s that satisfy Eq.~\ref{MinkN}.  This surface has infinite area, which relates to the  divergent number density of Minkowski vacua in Fig.~\ref{fig:BPHisto}.

Equation \ref{extremum}, together with the definition of $\widetilde{V}$ in Eq.~\ref{Vtildedef}, gives $N$ coupled quadratic equations for the $R_j^{-2}$s. B\'ezout's theorem then implies that there are generically $2^N$ solutions.  However, not all of these solutions need be physically relevant, because the $R_j^{-2}$s may be negative or complex.  (If $R_j^{-2}$ is negative, then the solution can be analytically continued $R_j\rightarrow i R_j, \theta_j\rightarrow i \theta_j$, so that the $j$th internal manifold is hyperbolic instead of spherical.)  For a given set of $n_j$, there are therefore at most $2^N$ physical critical points, though there may be fewer and, as we will see, there may even be none.

\subsection{Radion masses}
\label{sec:radionmass}
The Hessian matrix $H_{ij}^{(\phi)}$ is the matrix of second derivatives about an extremum in terms of the canonical $\phi_i$ coordinates. Its eigenvalues give the mass squareds of the radion fluctuations. To find $H_{ij}^{(\phi)}$, we will first compute  $H_{ij}^{(R)}$, the matrix of second derivatives about an extremum in terms of the non-canonical $R_i$ coordinates:
\begin{gather}
\label{HessNonCanon}
H^{(R)}_{ij}\equiv\frac{\partial^2 V}{\partial R_i\partial R_j}\Bigg|_\text{ext}=4V\left[\frac{1}{R_i^{\,2}}\left(\frac1{\widetilde V R_i^{\,2}}-2\right) \delta_{ij} - \frac{1}{R_i R_j}\right] = \frac{4V}{R_iR_j}\left[B_i\delta_{ij} - 1\right],
\end{gather}
where in the last equality we've defined the combination
\begin{gather} \label{FixedLabel3Jan08}
B_i\equiv\frac1{\widetilde V R_i^{\,2}}-2=-\,\frac{4n_i^{\,2}-R_i^{\,2}}{2n_i^{\,2}-R_i^{\,2}} = \frac{1}{4} \frac{\partial^2 \log \widetilde{V}}{(\partial \log R_i)^2} .
\end{gather}
Because we are expanding around an extremum, the conversion to canonical coordinates is given by multiplying by two factors of the Jacobian matrix
\begin{gather}
H^{(\phi)}_{ij}\equiv\frac{\partial^2 V}{\partial \phi_i\partial \phi_j}\Bigg|_\text{ext}=\sum_{k=1}^N\sum_{l=1}^N \frac{\partial R_k}{\partial\phi_i} \frac{\partial R_l}{\partial\phi_j} H^{(R)}_{kl}=2V\sum_{k=1}^N\sum_{l=1}^N (\delta_{ik}-A)(\delta_{jl}-A)(B_k\delta_{kl}-1) \nonumber\\
=2V\left[B_i\delta_{ij}-1-A(B_i+B_j)+2AN-N^2A^2 + A^2 \sum_{k=1}^NB_k\right].
\label{Hessphi}
\end{gather}
The trace of this mass matrix gives the sum of the mass squareds
\begin{equation}
\textrm{tr} H^{(\phi)}_{ij}=\frac{2NV}{N+1}\left(-1+\sum_{k=1}^N B_k\right). 
\label{tracephi}
\end{equation}
The determinant of this mass matrix gives the products of the mass-squarers and is itself given by the product of the determinants of its factors
\begin{equation}
\label{Hphidet}
\det H^{(\phi)}_{ij}=\frac1{2^N}\left[\det(\delta_{ij}-A)\right]^2 \det (R_kR_l H^{(R)}_{kl})=\frac1{N+1}\left(\prod_{k=1}^N 2VB_k\right)\left(1-\sum_{k=1}^N\frac1{B_k}\right).
\end{equation}

\subsection{Catastrophes}
\label{sec:catastrophes}

Catastrophes are the surfaces in $n_i$-space across which the number of physical extrema of the potential changes.  Varying the $n_i$ can cause two extrema (a saddle point with $k$ negative modes and a saddle point with $k\pm1$ negative modes) to merge, annihilate, and move off into the complex plane as conjugate pairs (so that two $R_i^{\, 2}$s become complex), or it can cause a single extremum to run away to infinity (so that an $R_i^{-2}$ becomes negative). At a catastrophe, the matrix of second derivatives  has a zero-mode along the direction joining the merging critical points, or if it is going to infinity along the direction joining the extremum to infinity. A catastrophe therefore occurs whenever Eq.~\ref{Hphidet} crosses zero,
\begin{gather}
\label{catastrophe}
\text{catastrophe:}\hspace{.5in}\sum_{i=1}^N \frac1{B_i}=1.
\end{gather}

\subsection{Minima of the effective potential}
\label{sec:minimum}

We are most interested in the minima of the effective potential, for which the canonically normalized Hessian $H_{ij}^{(\phi)}$ has only positive eigenvalues.

All AdS extrema are minima.  ($H_{ij}^{(\phi)}$ is positive-definite if and only if the matrix $R_iR_jH_{ij}^{(R)}=V[B_i\delta_{ij}-1]$ is, and we can use Sylvester's criterion, which states that a symmetric matrix is positive-definite when all of its principle minors are positive-definite.  AdS extrema have both $V<0$ and $B_i<-2$, which is sufficient to prove that $H_{ij}^{(\phi)}$ is positive-definite.) On the other hand, de Sitter extrema can have any number of negative modes. 

Equation~\ref{extremum} has $N$ choices of $\pm$ sign, one for each $R_i$. We will now show that  minima necessarily lie in the all-minus branch, so that the minimum is  the most compact of all the extrema:
\begin{equation}
R_i^{\,2} \biggl|_{\textrm{min}}  = \frac{1- \sqrt{1-8n_i^{\,2}\widetilde V}}{2\widetilde V}.  \label{NRminimum} 
\end{equation}
For dS minima, this can be established by arguing that it is only on the all-minus branch that $B_i>1$ for all $i$, a necessary condition for $R_iR_jH_{ij}^{(R)}$ to be positive-definite. For AdS minima, it follows from the fact that only the all-minus branch corresponds to $R_i^{2}>0$ for all $i$.
 Equation~\ref{NRminimum} implies that  at a minimum 
\begin{eqnarray}
 \widetilde{V} \biggl|_{\textrm{min}} & = & 1 - \sum_i \frac{1 + 4 n_i^2 \widetilde{V}  + \sqrt{1 - 8 n_i^2 \widetilde{V}} }{8 n_i^2} , \label{eq:Vintermediate}\\
 B_i \widetilde V \biggl|_{\textrm{min}} &=&  \frac{1 - 8n_i^2  \widetilde{V} + \sqrt{1 - 8 n_i^2 \widetilde{V}} }{4n_i^2}  . 
 \label{eq:doublederivativeVtilde}
\end{eqnarray}

\subsection{Dirac quantization condition}
\label{Dirac}
The number of flux units in this theory is quantized by a Dirac condition.  The integral of $F$ around an internal 2-sphere must be a discrete multiple of the magnetic flux quantum, $g$. The $n_i$ must satisfy
\begin{gather}
n_i=\frac{g}{4\pi} \sqrt{\frac{\Lambda_{D}}{2}}\,\mathbb{Z}, \label{eq:quantization}
\end{gather}
for somer integer $\mathbb{Z}$.

In this paper, we will often be interested in computing the number of flux vacua with a given property, for instance those with a cosmological constant that lies in a given range.  Following Ashok and Douglas \cite{Ashok:2003gk}, for much of this paper we will treat the $n_i$ as continuous variables, an approximation that's valid when $g\sqrt{\Lambda_D} \ll1$.  (This is the same approximation as was used to derive the histogram for the BP model in \cite{Bousso:2000xa} and in Eq.~\ref{BPHisto}.)  The continuous approximation allows us to replace sums over vacua with integrals over $n_i$, and to use the volume of $n_i$-space as a proxy for the number of vacua. For instance, the number density of vacua can be approximated as
\begin{equation}
\frac{d \hspace{.2mm} \text{Num}}{dV_4} \sim  \left( \frac{4\pi}{g} \sqrt{\frac2\Lambda} \right)^N \frac{d \hspace{.2mm} \text{Vol}_{\textrm{flux}}}{dV_4},
\end{equation}
where $\textrm{Vol}_\textrm{flux}$ is the volume in $n_i$-space 
\begin{equation}
\frac{d \hspace{.2mm} \text{Vol}_{\textrm{flux}}}{dV_4} \sim\frac1{dV_4} \int_{ V_4}^{ V_4 + dV_4} dn_1\cdots dn_N .
\label{continuousapprox}
\end{equation}

\section{$\bf{N=1}$}
\label{N1}
For $N=1$, there is a single internal 2-sphere of radius $R$, wrapped by flux.   This model was originally studied by Freund and Rubin in the case where the higher-dimensional $\Lambda_D=0$ \cite{Freund:1980xh}, and was soon generalized to $\Lambda_D\ne0$ \cite{RandjbarDaemi:1982hi}.  The landscape that arises is qualitatively different depending on the sign of $\Lambda_D$, as is summarized in Fig.~\ref{fig:TableN1}.  It will be helpful in this section to consider both signs of $\Lambda_D$, since the $N\ge2$ FR model (despite itself having $\Lambda_D>0$) has some regimes in which it behaves like the $N=1$ FR model with $\Lambda_D>0$ and other regimes in which it behaves like the $N=1$ FR model with $\Lambda_D<0$.  

 The effective potential is the $N=1$ version of Eq.~\ref{effpotRLambdaneg}:
\begin{gather}
V=\frac1{R^2}\left[\frac{n^2}{R^4}-\frac{1}{R^2}+\text{sign}({\Lambda_D})\right].
\end{gather}
We will calculate the minimum of this potential, and see that the scaling  changes when $n\sim 1$.

\begin{center}
\begin{tabular}{| c !{\vrule width 2pt} c   !{\vrule width 2pt}  c  !{\vrule width 2pt}  c |}
\hline
 & $n\ll\sqrt{1/3}$   & \multicolumn{2}{c|}{$n\gg\sqrt{1/3}$}  \T\B  \\ \cline{2-4}
 &\begin{tabular}{@{}c@{}} \hspace{1.5in} \\any $\Lambda_D$ \\\, \end{tabular}     &  \begin{tabular}{@{}c@{}} \hspace{1.5in} \\ $\Lambda_D<0$ \\ \hspace{1.5in} \end{tabular}     &  \begin{tabular}{@{}c@{}} \, \\ $\Lambda_D>0$ \\ \, \end{tabular}    \T\B \\ \noalign{\hrule height 2pt}
 \begin{tabular}{@{}c@{}} terms balanced \\ in the potential \end{tabular} &   flux and curvature  & flux and $\Lambda_D$  &  \hspace{1.5in} \T\B \\
  \cline{1-3}
$V_4$ & $ -\dfrac4{27}\dfrac1{n^4}$&$-\dfrac2{3\sqrt3}\dfrac1{n}$ & \T\B \\
\cline{1-3}
$\widetilde V_4$ & $-\dfrac29\dfrac1{n^2}$ & $-\dfrac23$ &  no solutions   \T\B \\ \cline{1-3}
$R^2$ & $\dfrac32n^2$&$\sqrt3n$&  \T\B \\ \cline{1-3}
$m^2$& $\dfrac89\dfrac1{n^4}$&$\dfrac2{\sqrt3}\dfrac1n$& \T\B \\
\hline
\end{tabular}
\end{center}
\vspace{-.19in}
\makebox[5.35in][r]{}\vspace{.11in}
\vspace{-.25in} 
\begin{figure}[htbp] 
\caption{A table summarizing the asymptotic behavior of the $N=1$ FR model for both signs of $\Lambda_D$.  At small $n$, the radius of the internal sphere is far smaller than the higher-dimensional Hubble length, and so $\Lambda_D$ is unimportant.  For large $n$, the behavior is dominated by the cosmological constant.  Intermediate $n$ exhibits transitional behavior; this is where, for $\Lambda_D>0$, there are de Sitter vacua.}
  \label{fig:TableN1}
\end{figure}

\begin{figure}[t] 
   \centering
   \includegraphics[width=\textwidth]{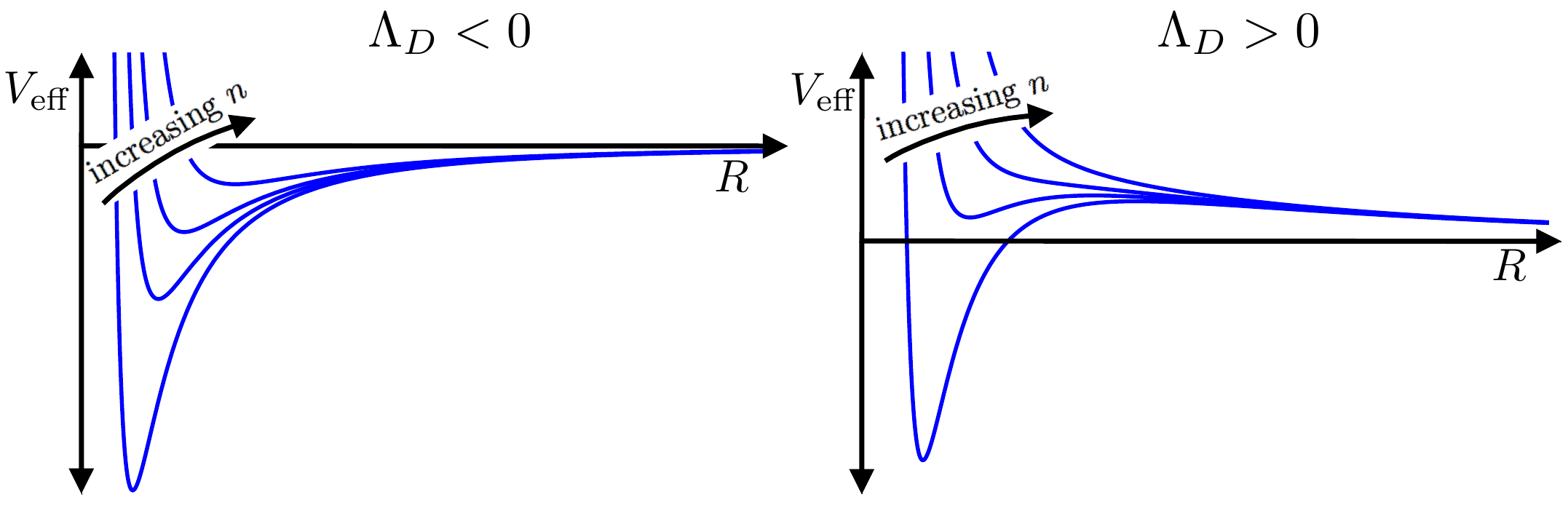}
   \caption{The Einstein-frame effective potential $V_{}$ for radius $R$ of the single internal 2-sphere in the $N=1$ case.  The effective potential has qualitatively different behavior depending on the sign of $\Lambda_D$.  If $\Lambda_D<0$, then the potential has a single AdS minimum for any number of flux lines $n$.  If $\Lambda_D>0$, then the potential has either two, one, or zero extrema depending on the magnitude of $n$; the maximum is always dS, while the minimum can be either AdS, Minkowksi, or dS. }
    \label{fig:effpotN1}
\end{figure}

\begin{figure}[t] 
   \centering
   \includegraphics[width=\textwidth]{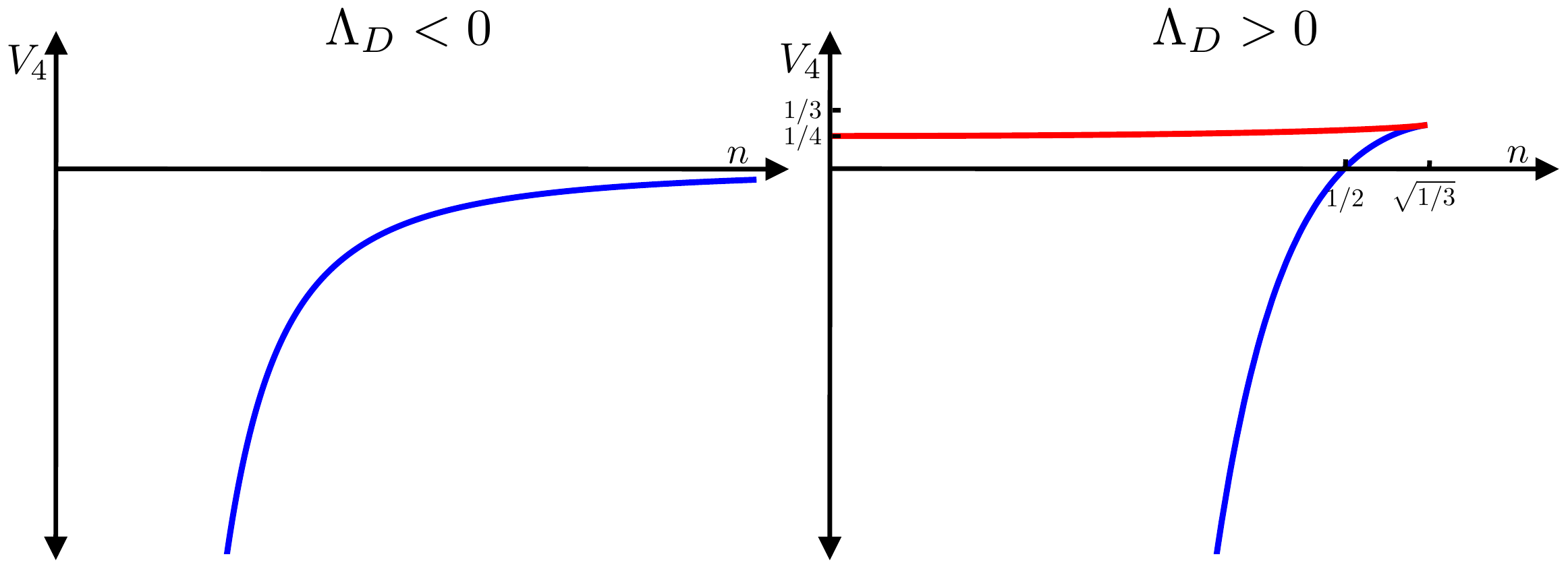} 
   \caption{The value $V_4$ of the effective potential at its extrema as a function of $n$.  Minima of the potential are shown in blue and maxima are shown in red.  If $\Lambda_D<0$, there is a single AdS minimum for every value of $n$.  If $\Lambda_D>0$, there is a minimum and a maximum when $n<\sqrt{1/3}$ and no extrema when $n>\sqrt{1/3}$.  The maximum is always dS; the minimum is AdS when $0<n<1/2$ and dS when $1/2<n<\sqrt{1/3}$.}
   \label{fig:N1V4vn}
\end{figure}

\begin{figure}[t] 
   \centering
   \includegraphics[width=\textwidth]{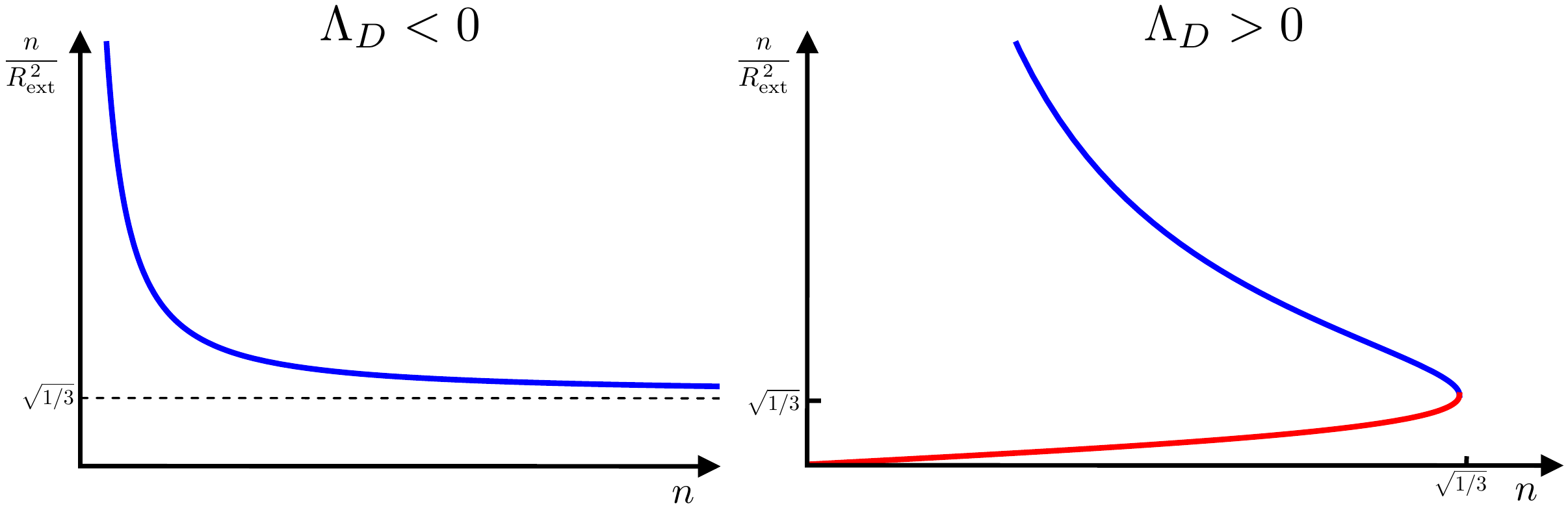}
   \caption{The density of flux lines at an extremum $n/R_\text{ext}^{\,2}$ as a function of $n$; minima are shown in blue and maxima are shown in red. Unlike the BP model, the density of flux lines in a minimum is a falling function of the number of flux lines. If $\Lambda_D<0$, there is a single minimum whose internal volume $R_\text{min}^{\,2}$ grows fast enough with $n$ that $n/R_\text{min}^{\,2}$ is monotonically decreasing with $n$; the flux density asymptotes to a constant for the vacua with large $n$ and large internal volume.  If $\Lambda_D>0$, the flux density is a falling function of $n$ for the minimum and a growing function of $n$ for the maximum.}  \label{N1csaregoingdowndowndown}
\end{figure}

\begin{figure}[t] 
   \centering
   \includegraphics[width=\textwidth]{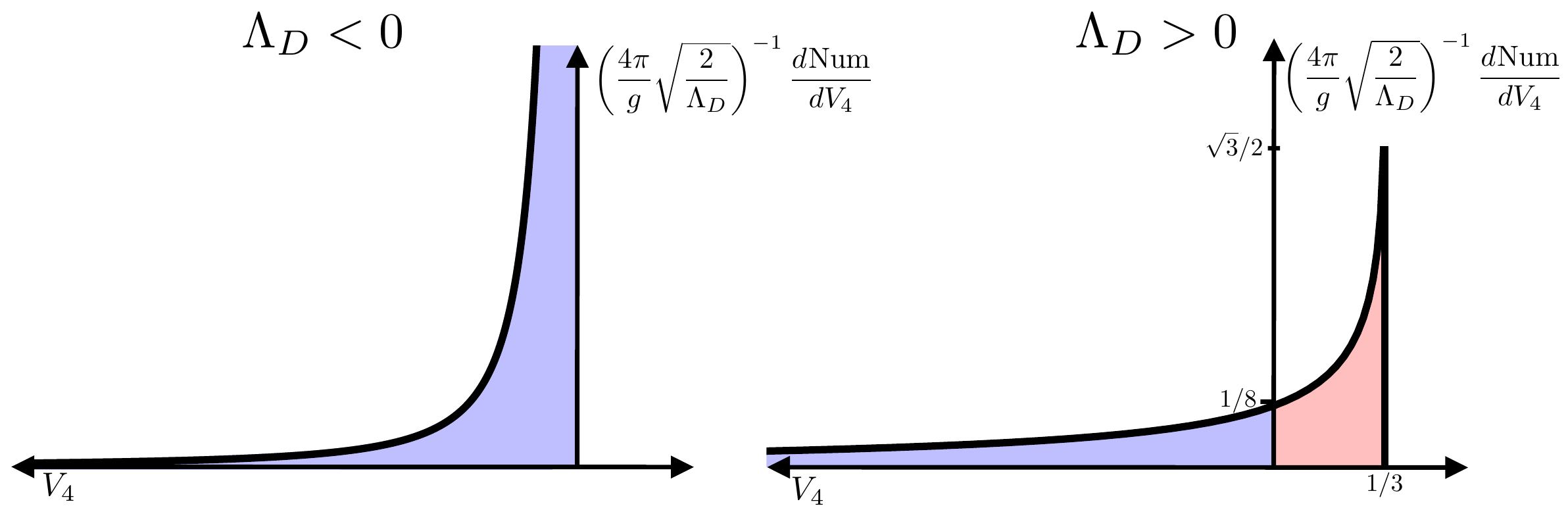} 
   \caption{A histogram of the number of vacua with a given value of $V_4$.  If $\Lambda_D<0$, there is an accumulation point of vacua with $V_4$ just below 0.  If $\Lambda_D>0$, there are a finite number of AdS vacua and a finite number of dS vacua.  The distribution is smooth through $V_4=0$, like in the BP landscape.  One difference from the BP landscape is that the distribution cuts off at $V_4=1/3$ not because effective field theory is breaking down, but instead because of decompactification---the minimum and the maximum of the potential have annihilated in a fold catastrophe. 
   }   \label{fig:N1Histo}
\end{figure}

\subsection{$\mathbf{\Lambda_D<0}$}
\label{sec:N1negative}
The effective potential for $\Lambda_D < 0$ is plotted in the left panel of Fig.~\ref{fig:effpotN1}. There is only ever a single extremum of the potential, and it is always an AdS minimum:
\begin{gather}
\label{N1Lambdanegmin}
R^2=-1+\sqrt{1+3n^2}, \hspace{.3in}\text{and}\hspace{.3in}
V_4=-\frac{1+2n^2-\sqrt{1+3n^2}}{R^6}. 
\end{gather}
The left panel of Fig.~\ref{fig:N1V4vn} shows the value of the potential in this minimum, $V_4$,  as a function of $n$.
The radion mass is given by the second derivative about the minimum, not with respect to $R$, but with respect to the canonical coordinate $\phi= 2 \log R$:
\begin{gather}
m^2=\frac{2\sqrt{1+3n^2} }{R^4}.
\end{gather}

The flux is a repulsive term in the effective potential  that buttresses the extra dimensions against collapse and dominates at the smallest values of $R$. Since $\Lambda_D <0$, the other terms in the effective potential are both attractive, and want to make the extra dimensions shrink. The curvature term falls off as $R^{-4}$, whereas the cosmological constant term falls off as $R^{-2}$. At small $R$ ($n \ll 1$) the minimum is determined by the interplay of flux and curvature; at large $R$ ($n \gg 1$) the minimum is determined by the interplay of flux and $\Lambda_D$. The different scaling in these two limits are shown in Fig.~\ref{fig:TableN1}. 

Unlike in the BP model, we can raise $n$ arbitrarily large without the energy density becoming super-Planckian. Increasing $n$ increases the number of flux lines, but it also shifts the minimum out to larger volumes, and this shift more than compensates for the increased number of lines.  As a consequence the higher-dimensional energy density in the minimum is now a falling function of $n$, as shown in the left panel of Fig.~\ref{N1csaregoingdowndowndown}.     

The distribution of minima as a function of $V_4$ is plotted in the left panel of Fig.~\ref{fig:N1Histo}. In the continuous approximation of Eq.~\ref{continuousapprox} the number of vacua is proportional to the length in $n$-space. Because most of the vacua lie at large $n$ and have small $V_4$, we can Taylor expand 
\begin{gather}
\frac{d\hspace{.2mm}\text{Num}}{dV_4} \sim \frac{4\pi}{g} \sqrt{\frac{2}{\Lambda_D}} \frac{dn}{dV_4} =
\begin{cases} \frac{4\pi}{g} \sqrt{\frac{2}{\Lambda_D}}\frac{2}{3\sqrt3} \frac1{ V_4^{\,2}} + O\left(\frac1{ V_4}\right), & \text{if }V_4<0\\[.25in] 0,& \text{if }V_4\ge0. \end{cases}
\end{gather}
The singularity is non-integrable and the infinity of AdS vacua accumulate at $V_4 = 0^-$.

\subsection{$\mathbf{\Lambda_D>0}$}
\label{sec:N1positive}
The effective potential for $\Lambda_D > 0$ is plotted in the right panel of Fig.~\ref{fig:effpotN1}. The potential has  two extrema
\begin{gather}
R^2=1\pm\sqrt{1-3n^2},\hspace{.3in}\text{and}\hspace{.3in}
V_4=\frac{1-2n^2\pm \sqrt{1-3n^2}}{R^6}. \label{eq:N1positiveLambda}
\end{gather}
When $n<\sqrt{1/3}$, the minus branch corresponds to a minimum and the plus branch corresponds to a maximum; when $n>\sqrt{1/3}$, the extrema are complex conjugate pairs and do not correspond to physical solutions. The right panel of Fig.~\ref{fig:N1V4vn} shows both $V_4$'s as a function of $n$.  The maximum and minimum satisfy
\begin{equation}
R_\text{min}^{\,2} + R_\text{max}^{\,2}=2,\hspace{.3in}\text{and}\hspace{.3in} R_\text{min}^{\,2} R_\text{max}^{\,2}=3 n^2. \label{eq:N1Rrelations}
\end{equation}
The radion mass at an extremum is
\begin{gather} 
m^2=\mp\frac{2\sqrt{1-3n^2} }{R^4},
\end{gather}
which indeed is negative for the maximum and positive for the minimum. As  $n \rightarrow \sqrt{1/3}$, the catastrophe condition Eq.~\ref{catastrophe} is satisfied and the masses converge at zero.  This extra zero-mode is the indication that the minimum and maximum are merging at an inflection point and moving off into the complex plane.      

As in the $\Lambda_D < 0$ case the flux is a hard-core repulsion term in the effective potential and the curvature is a medium-range attractive term. The difference is that the long-range term given by $\Lambda_D$ is now repulsive. This means there are now no solutions with large $R$ because they get swept out by the Hubble expansion towards decompactification ($R \rightarrow \infty$).  For $N=1$ and $\Lambda_D > 0$, there are no large-flux minima and the landscape is finite.

On the other hand, for solutions with small $R$ ($n \ll 1$) the size of the minima is determined by the interplay of flux and curvature and the cosmological constant is irrelevant. Indeed the small $n$ behavior of the $\Lambda_D >0 $ is the same as that for $\Lambda_D<0$. This means that all small $n$ vacua are AdS. 

It is at intermediate $n$ that the de Sitter minima live. When $n=1/2$ the Minkowski condition of Eq.~\ref{MinkN} is satisfied and $V_4 = 0$ at the minimum. When $1/2<n<\sqrt{1/3}$ the minimum is de Sitter. For these de Sitter minima the contributions of flux, curvature and cosmological constant are roughly equal.

The distribution of minima as a function of $V_4$ is plotted in the right panel of Fig.~\ref{fig:N1Histo}. There are three differences from the $\Lambda_D<0$ case. First, since decompactification puts a bound on $n$ there are now only a finite number of minima. Second, and relatedly,  the distribution of vacua is now smooth through $V_4 = 0$---the accumulation point has vanished.  Third, there are now de Sitter vacua in the landscape; indeed these compose an O(1) fraction of the vacua: a fraction $\sqrt3/2$ are AdS and a fraction $1-\sqrt{3}/2$ are de Sitter. 

The histogram for the $N=1$ theory with  $\Lambda_D>0$   resembles the histogram for the BP model (in the left panel of Fig.~\ref{fig:BPHisto}), but there is an important difference.  Like the $\Lambda_D\le0$ landscape and unlike the BP landscape, the higher-dimensional energy density in the minimum is uniformly falling in $n$.  The distribution cuts off sharply at $n=\sqrt{1/3}$ not because the energy densities are becoming Planckian, but because raising $n$ causes the minimum and maximum to annihilate.

\subsection{$\bm{ \Lambda_D=0$}}
\label{sec:N1zero}
The $n \ll 1$ limits of both the $\Lambda_D > 0$ and the $\Lambda_D < 0$ theories agree. The reason they agree is that the cosmological constant is unimportant in this limit---the flux and curvature terms in the action dominate ($F^2 \sim R^{-2} \gg \Lambda_D$). The $\Lambda_D = 0$ theory, therefore, is given precisely by this limiting behavior, and the answers in the right column of Fig.~\ref{fig:TableN1} are exact. (Though the definitions of $R$ and $n$ are rescaled by factors of $\Lambda_D$ as given in Eqs.~\ref{metricansatz} and \ref{fluxconfiguration}, in the calculation of physical quantities these dependencies cancel in the $\Lambda_D \rightarrow 0$ limit.)  

Since there is no cosmological constant to contend with, $V_4$ tends more rapidly to $0^{-}$ as $n \rightarrow \infty$, as is reflected in the distribution of vacua
\begin{gather}
\frac{d\hspace{.2mm}\text{Num}}{dV_4}  \sim \frac{4\pi}{g} \sqrt{\frac{2}{\Lambda_D}} \frac{dn}{dV_4} =\begin{cases} \frac{4\pi}{g} \sqrt{\frac{2}{\Lambda_D}}\frac1{1728^{1/4}} \frac1{ V_4^{\,5/4}} & \text{if }V_4<0\\[.25in] 0 & \text{if }V_4\ge0. \end{cases}
\end{gather}

\section{$\mathbf{N=2}$}
\label{N2}

We will now study the case of 2 two-spheres, which is the simplest case rich enough to exhibit the new phenomena. $N=2$ has the additional advantage that it is easy to visually depict because the number of cycles coincides with the dimensionality of this page.

\subsection{Effective potential}
For $N=2$, the effective potential (Eq.~\ref{effpotR}) is a function of  the two radii $R_1$ and $R_2$, 
\begin{equation}
V_{}(R_1,R_2)= \frac{1}{R_1^{\,2}} \frac{1}{R_2^{\,2}} \left[ \frac{n_1^{\,2}}{R_1^{\,4}} + \frac{n_2^{\,2}}{R_2^{\,4}} -\frac{1}{R_1^{\,2}}  -\frac{1}{R_2^{\,2}}  + 1 \right],
\end{equation}
where $n_1$ and $n_2$ give the flux wrapping the two-spheres and the higher-dimensional cosmological constant is taken to be positive here and for the rest of this paper. The potential goes to infinity when $R_1$ or $R_2 \rightarrow 0$, and goes to zero when $R_1$ or $R_2 \rightarrow \infty$. We will be interested in the critical points of this potential, and especially the minimum. 

\begin{figure}[h!] 
   \centering
      \includegraphics[width=\textwidth]{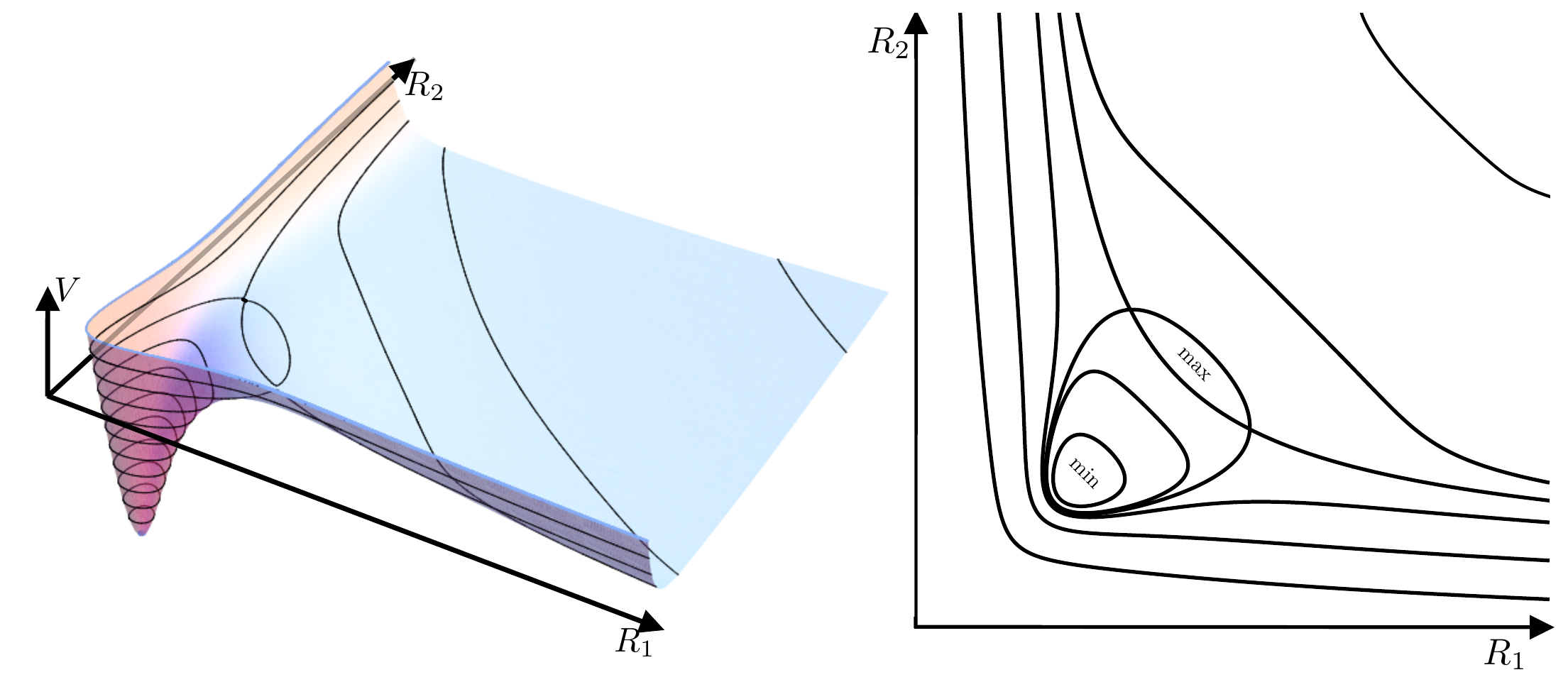}
\caption{Two depictions of the effective potential for $n_1 = n_2 = 0.68$. For this value of $n_1$ and $n_2$ the effective potential has 4 critical points, the most allowed by B\'ezout's theorem. The maximum and both saddles have $V_4 > 0$, and the minimum in this case has $V_4<0$. The potential goes to infinity as $R_1$ or $R_2 \rightarrow 0$, and to zero as $R_1$ or $R_2 \rightarrow \infty$. Because for this plot we have chosen $n_1 = n_2$, the potential has an $R_1 \leftrightarrow R_2$ symmetry. }
\label{3Dpotential}
\end{figure}

Figure~\ref{3Dpotential} shows the effective potential for a particular value of $n_1$ and $n_2$. For this value, there is ${2 \choose 0} = 1$ minimum, ${2 \choose 1} = 2$ saddle points, and ${2 \choose 2} = 1$ maximum. We showed in Sec.~\ref{extremaequations} that there are exactly $2^2 =4$ (possibly complex) critical points of the $N=2$ potential, so for the values of $n_1$ and $n_2$ shown in Fig.~\ref{3Dpotential} all of the critical points correspond to real physical solutions with $R_i^{-2}>0$. However, for other values of $n_1$ and $n_2$ not all critical points are physical.

\subsection{The phase diagram}
\begin{figure}[h!] 
   \centering
   \includegraphics[width=.83\textwidth]{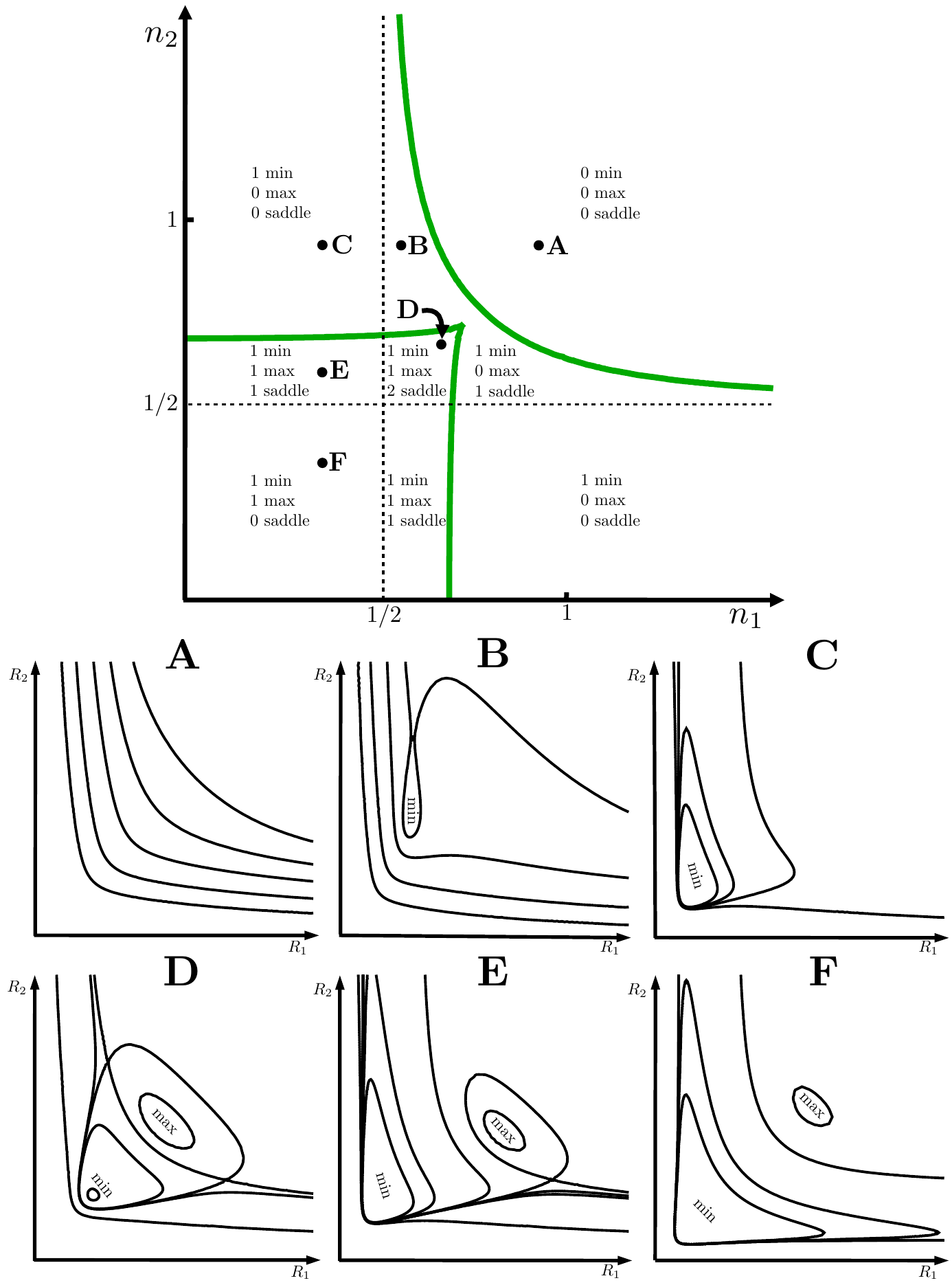}
   \caption{The top panel shows the catastrophes of the $N=2$ theory in the $(n_1,n_2)$-plane.  Crossing a catastrophe line changes the number of extrema; a green line causes a saddle to merge with either a min or a max, and a dotted black line causes a saddle to travel off to infinity.  The point where the two green fold-catastrophes lines merge is a cusp catastrophe.  Six sample points are chosen, each with a different distribution of extrema, and topological maps of $V_{}(R_1,R_2)$ are shown.  When $R_1$ and $R_2$ are both large, the potential tends to $V_\text{eff}=0$ from above; whenever either $R_1$ or $R_2$ goes to zero, the potential goes to $V_\text{eff}\rightarrow+\infty$.}
   \label{fig:ToposN2}
\end{figure}

Figure~\ref{fig:ToposN2} shows, for each value of $n_1$ and $n_2$, the number of physical critical points of the effective potential. 
This number changes along catastrophes; in the $N=2$ case the formula for the catastrophes Eq.~\ref{catastrophe} becomes 
\begin{gather}
\label{catastropheN21}
n_1=\frac12,
\end{gather}
or
\begin{gather}
\label{catastropheN22}
n_2=\frac12,
\end{gather}
or
\begin{gather}
65536 n_1^{\,6} n_2^{\,6}-67584 (n_1^{\,4} n_2^{\,6} + n_1^{\,6} n_2^{\,4}) +22800 (n_1^{\,6} n_2^{\,2}+ n_1^{\,2}
n_2^{\,6})-15420 (n_1^{\,2} n_2^{\,4} + n_1^{\,4} n_2^{\,2}) \nonumber \\
 +2358 n_1^{\,2} n_2^{\,2} - 2500 (n_1^{\,6}+ n_2^{\,6})  + 58848 n_1^{\,4} n_2^{\,4}+1125 (n_1^{\,4} + n_2^{\,4})
=0.
\label{catastropheN23}
\end{gather}
Fans of the binary system will recall that $65536 = 2^{2^{2^2}}$. 

Equations \ref{catastropheN21} and \ref{catastropheN22}  give the  dotted black lines in Fig.~\ref{fig:ToposN2}. Crossing these lines changes the number of saddles by one.  As $n_1 \rightarrow \frac{1}{2}$ from above, the $R_2^{-2}$ of a saddle goes through zero. For $n_1 < \frac{1}{2}$ the saddle has moved out to infinity and is no longer a physical solution. 

 Equation \ref{catastropheN23} has two branches, which give the green lines in Fig.~\ref{fig:ToposN2}. Crossing these lines changes the number of real critical points by two: across these lines two critical points merge and move off into the complex plane. Along the lower branch, the maximum merges with a saddle, so the potential has a maximum only beneath this line.  Along the upper branch, the minimum merges with a saddle, so the potential has a minimum only beneath this line.

Together, these  catastrophes divide the $(n_1,n_2)$-plane into regions with different numbers of critical points; the effective potential is drawn for a representative value of each region in the lower panels of Fig.~\ref{fig:ToposN2}.

\subsection{de Sitter and AdS minima}
\begin{figure}[t] 
   \centering
   \includegraphics[width=.6\textwidth]{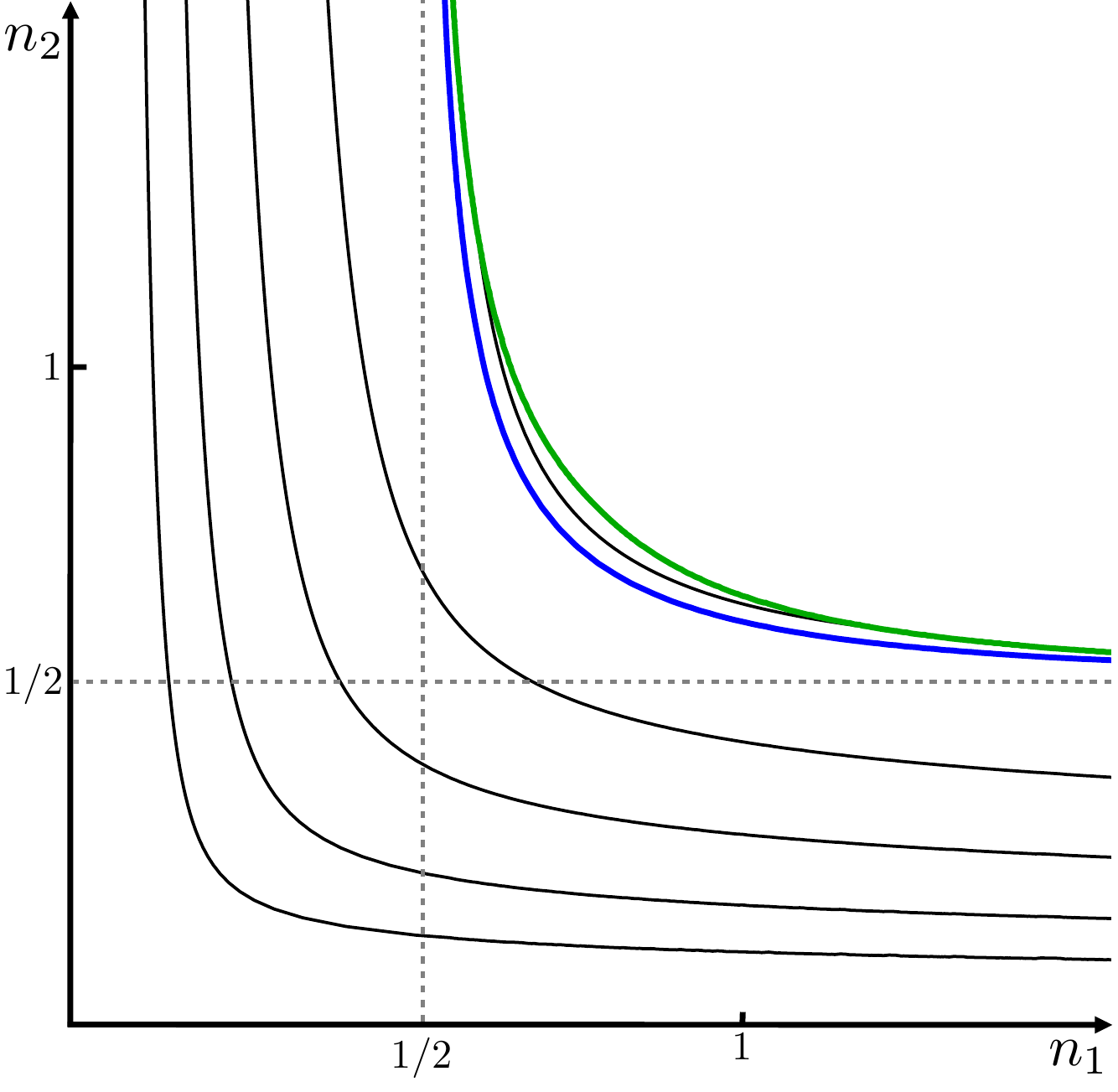}
   \caption{A plot showing equipotentials $V_4$ at the minimum, as a function of $n_1$ and $n_2$.  In blue is the Minkowski equipotential, with $V_4=0$; it asymptotes $n_1=1/2$ and $n_2=1/2$.  Equipotentials that lie below it correspond to AdS values; plotted are $V_4=-2.5$, $-25$, $-250$, and $-2500$.  Each AdS equipotential is infinitely long and asymptotes to $n_1=0$ and $n_2=0$.  Equipotentials that lie above the blue line correspond to de Sitter values, plotted is $V_4=+.025$.  De Sitter equipotentials are finite in length, they end abruptly when they intersect the green catastrophe line. }
   \label{EquipotentialsN2}
\end{figure}

There is at most one local minimum of the effective potential. We saw in Eq.~\ref{MinkN} that when $n_1$ and $n_2$ satisfy
\begin{gather}
\label{Mink2}
\text{Minkowski:}\hspace{.5in}\frac1{n_1^{\,2}}+\frac1{n_2^{\,2}}=4,
\end{gather}
this minimum has $V_4 = 0$. For lesser flux values the minimum has $V_4 < 0$; for greater flux values the minimum, if it exists, has $V_4 > 0$. Lines in $(n_1,n_2)$-space along which the minimum of the effective potential has the same value of $V_4$ are plotted in Fig.~\ref{EquipotentialsN2}.

The lines that correspond to $V_4<0$ are infinitely long and asymptote to $n_1=0$ and $n_2=0$. The line that corresponds to $V_4=0$ is infinitely long and asymptotes to $n_1=1/2$ and $n_2=1/2$.  The lines that correspond to $V_4 > 0$ are finite in length, ending abruptly when they intersect the catastrophe line.  The highest de Sitter minimum in the landscape has $n_1=n_2=3/4$ and $V_4=2/27$, saturating the catastrophe condition Eq.~\ref{catastropheN23}.   The lowest AdS vacua in the landscape have $n_1=0$ or $n_2=0$ and $V_4\rightarrow-\infty$ \cite{BAMBAMBAM}.

\subsection{Sequential compactification and the large-flux limit: $\mathbf{n_2 \gg n_1}$}
\label{subsec:sequential}
 
The region of the phase diagram with $n_2 \gg n_1$ can be understood from the perspective of sequential compactification. When  a solution  has $R_2 \gg R_1$  it is  a good approximation to think of the $N=2$ compactification as two, sequential, $N=1$ compactifications. First we compactify from eight to six dimensions on the small sphere of radius $R_1$; this gives rise to an effective six-dimensional cosmological constant $\Lambda_6$. Then we do a second $N=1$ compactification on the large sphere of radius $R_2$ to get down from six to four dimensions. The hierarchy of scales between the compactifications ensures the second compactification doesn't backreact on the first. (When $n_2 \sim n_1$ the sequential compactification analysis is not useful because  of backreaction.)

This explains the importance of $n_1 = \frac{1}{2}$. For $n_1$ less than $\frac{1}{2}$ the first $N=1$ compactification gives rise to a minimum with negative $\Lambda_6$. The second $N=1$ compactification is thus a $\Lambda_6 < 0$  compactification, which we saw in Sec.~\ref{sec:N1negative} always gives rise to exactly one critical point: an AdS minimum. Correspondingly the region with $n_1 < \frac{1}{2}$ and $n_2$ huge has an AdS minimum, and indeed in this parameter regime the AdS minimum is the only critical point.\footnote{The first $N=1$ compactification from eight to six dimensions, as well as giving rise to a 6d AdS minimum, also gives rise to a 6d dS maximum, as we saw in Sec.~\ref{sec:N1positive}. However, that 6d dS maximum does not in turn give rise to any physical 4d critical points in the limit of huge $n_2$, because, as we also saw in Sec.~\ref{sec:N1positive}, large enough flux  makes de Sitter solutions decompactify.} Just as in the $N=1$ case with $\Lambda_6 \leq0$, there is a path that connects this minimum to infinity along which the potential is never positive. 
 
 On the other side of the line, for $n_1$ just larger than $\frac{1}{2}$, the first $N=1$ compactification gives rise to a minimum with positive $\Lambda_6$. The second $N=1$ compactification is thus a $\Lambda_6 > 0$ compactification, which as we saw in Sec.~\ref{sec:N1positive} gives rise to either zero or two critical points: when the flux is small enough, there is an AdS minimum and an unstable dS solution; as the flux  is raised, the minimum first moves from AdS to dS and then, if the flux is raised further, merges with the unstable solution and disappears. Correspondingly the region with $n_1$ just bigger than a half and $n_2$ huge  may have an AdS minimum, a dS minimum or no minimum depending on $n_2$. 
As $n_1 \rightarrow \frac{1}{2}$ from above, $\Lambda_6 \rightarrow 0$ from above. 
This means that both the Minkowski line and the catastrophe line move off to $n_2 = \infty$, as they would for an $N=1$ compactification with vanishing cosmological constant, and that the solution with one negative mode grows to infinite size.

In the next section we will make this approach quantitative and use it to derive the sizes of the extra dimensions and the value of the potential in the minimum in the large-flux limit not just for $N=2$ but for general $N$. 

\section{General $\mathbf{N}$}
\label{sec:GeneralN}
For general $N$, the effective potential is a function of the $N$ radii $R_1, R_2, \cdots,$ and $R_N$, and is given by Eq.~\ref{Vtildedef} as
\begin{equation}
V_{}(R_i)=\left\{\prod_{i=1}^N\left(\frac{1}{R_i^{\,2}}\right)\right\} \widetilde V(R_i), \hspace{.3in} \text{where} \hspace{.3in}  \widetilde V(R_i) \equiv \sum_{i=1}^N \frac{n_i^{\,2}}{R_i^{\,4}}-\sum_{i=1}^N\frac1{R_i^{\,2}} + 1. \label{eq:beginningofsection6}
\end{equation}
In this section we will be interested in the critical points, and  especially the minimum, of this effective potential. For this minimum we will calculate  the sizes of the extra dimensions, the value of the potential, and the mass spectrum of small perturbations. 

Expressions for these quantities are given by solutions to the $N$ coupled quadratic equations of Eq.~\ref{extremum}, and the general solutions are therefore monstrous and unilluminating. 
However, we will see that there are two complementary limits  in which the answers simplify considerably.

The first limit is when all of the $n_i$'s are equal. In Sec.~\ref{subsec:generalNdiagonal} we will look at these permutation-symmetric  minima. These vacua are the least quintessentially general-$N$, in the sense that their behavior is most reminiscent of the $N=1$ Freund Rubin compactifications of Sec.~\ref{N1}.

The second limit is the large-flux regime. When some subset of the extra dimensions are much larger than the others, then, as we saw in Sec.~\ref{subsec:sequential}, their backreaction on the other extra dimensions can be safely neglected. In Sec.~\ref{subsec:GeneralNMinkowski} we will look at large-flux Minkowski minima; in Sec.~\ref{subsec:generalNdStail} we will look at large-flux de Sitter minima; and in Sec.~\ref{subsec:generalNnohumpAdS} we will look at large-flux AdS minima.

These two limits are complementary in the sense that one covers small fluxes and the other large fluxes, and together they cover the full range of qualitative behaviors exhibited by our model. We will see that it is possible to take hybrid limits that involve both hierarchies between the fluxes and some fluxes being equal to each other. In Sec.~\ref{sec:statistics} we will explain our interest in the large-flux limit by showing that large-flux minima constitute the great majority of all flux minima.

\subsection{General $\mathbf{N}$ phase diagram} 

For general $N$, the phase diagram of critical points is qualitatively similar to the $N=2$ case shown in Fig.~\ref{fig:ToposN2}. As before, there is only ever at most one minimum of the potential, but there are now more saddle points, as described in Sec.~\ref{extremaequations}, and their disappearance structure is more intricate. As in the $N=2$ case, a saddle point runs away to infinity and disappears whenever an $n_i^{-2}$ passes $4$, but now also saddle points reach infinity whenever there is some proper subset of the $n_i$ such that 
\begin{equation}
{\sum_{\textrm{proper subset}} \frac{1}{n_i^{\, 2}} = 4}. \label{eq:propersubset}
\end{equation}

This can be understood from the perspective of sequential compactification. As $n_i^{-2}\rightarrow 4^{-}$, an $N=1$ compactification on just the $i$th two-sphere gives  $\Lambda_{D-2} \rightarrow 0^{+}$, and the corresponding saddle points swell infinitely large. As $n_i^{-2} + n_j^{-2} \rightarrow 4^{-}$,  an $N=1$ compactification on either $R_i$ or $R_j$ individually still gives rise to a positive $\Lambda_{D-2}$, but an $N=2$ compactification on both simultaneously gives rise to $\Lambda_{D-4} \rightarrow 0^{+}$, and the corresponding saddle points run away to infinity. In general a saddle disappears to infinity whenever some $\Lambda_{D - 2k} \rightarrow 0^+$. 
 
Likewise, there are finite catastrophes with complicated formulae, the higher-dimensional generalizations of Eq.~\ref{catastropheN23}, but the only one of these catastrophes that affects the minimum is when a de Sitter minimum merges with a one-negative-mode saddle and disappears.

\subsection{Along the diagonal $\mathbf{n_1 = n_2 = \cdots =  n_N \equiv n}$}
\label{subsec:generalNdiagonal}

The novelty of  $N\geq2$ is to be found in the large-flux limit, which does not exist for $\Lambda_D > 0$ when $N=1$. To probe the more mundane behavior at moderate flux values, let's look at the case where all the flux values are the same. The minimum is then permutation-symmetric  with $R_1 = R_2 = \cdots = R_N \equiv R$, and we will be able to make use of the enhanced symmetry.

\subsubsection{The symmetric extrema}
The symmetric extrema satisfy a version of Eq.~\ref{extremum}
\begin{gather}
\frac{R^{\,2} - 2n^{2}}{R^{\,4}}  = \widetilde V(R_i)=\frac{Nn^2}{R^4}-\frac{N}{R^2}+1,
\end{gather}
which can be solved for $R$ to give
\begin{equation}
{R^2 = \frac{1}{2} \left( 1 + N \pm \sqrt{ \left(  1+ N  \right)^2 - 4 n^2 (2+N) } \right) }.
\end{equation}
The `minus' branch is the minimum, the `plus' branch is a saddle/maximum.  The corresponding value of the potential is
\begin{equation}
V = \frac{1}{2R^{2N + 4} }  \left( 1 + N - 4n^2 \pm \sqrt{ \left(  1+ N  \right)^2 - 4 n^2 (2+N) } \right).
\end{equation}
Larger $n$ means larger $R_\textrm{min}$ but smaller $R_\textrm{max}$; we have 
\begin{equation}
R^{\, 2}_\textrm{min} + R^{\, 2}_\textrm{max} = 1 + N \ \  \textrm{ and } \ \ 
R_\textrm{min}^{\, 2} R_\textrm{max}^{\, 2} = n^2(2+N).
\end{equation} 
For $N=1$ these expressions reduce to Eq.~\ref{eq:N1Rrelations}.\\

\subsubsection{The eigenvalues}
The canonically normalized Hessian matrix at an extremum Eq.~\ref{Hessphi} reduces to
\begin{gather}
H^{(\phi)}_{ij}=2V\left[B\delta_{ij}-1-2AB+2AN-N^2A^2 + N A^2B\right],
\end{gather}
where $B_1 = B_2 = \cdots = B = -(4n^2 - R^2)/(2n^2 - R^2)$. The eigenvectors of this Hessian are categorized by how they transform under the   permutation symmetry between the $N$ two-spheres.  One of the eigenvectors is the fluctuation in which all of the $\Delta \phi_i$ are equal; this preserves the symmetry and changes the total internal volume of the compactification.  The eigenvalue associated with this eigenvector is the smallest of all the eigenvalues and is given by 
\begin{gather}
m_{\textrm{sym}}^{\, 2}=2V[B +N(-1-2AB+2AN-N^2A^2+NA^2B)] = \frac{2 ( 4 n^2 - R^2 - \frac{2N}{1+N} n^2) }{R^{2N+4}}.
\end{gather}
The other $N-1$ eigenvectors break the permutation symmetry, but keep the total internal volume fixed. For example, $\Delta \phi_1 - \Delta \phi_2$ is an eigenvector that increases $R_1$ and decreases $R_2$ while keeping $R_1 R_2$ unchanged. 
The eigenvalue associated with these asymmetric  eigenvectors is
\begin{gather}
m_{\textrm{asym}}^{\, 2} =2VB= \frac{2 ( 4 n^2 - R^2)}{R^{2N+4}}.
\end{gather}
The symmetric mode is stable for the minimum and unstable for the saddle/maximum; it is zero at the fold catastrophe where the minimum and saddle merge. The asymmetric mode is stable for the entire minimum branch and for some of the maximum branch: on the `plus' branch $m_\textrm{asym}^{\, 2}$ is negative for $n^2 < \frac{2 + 3N}{16}$ and positive for $n^2> \frac{2 + 3N}{16}$.  The transition is a degenerate cusp catastrophe in which the symmetric 1-saddle spits off $N-1$ asymmetric 1-saddles and correspondingly picks up $N-1$ extra negative modes, becoming a maximum. For the case $N=2$ the value $n^2 = \frac{2 + 3N}{16} = \frac{1}{2}$ gives the apex of the lower catastrophe line shown in the top panel of Fig.~\ref{fig:ToposN2}.

\subsubsection{Special values of ${n}$}
There are three special values of $n$ at which the character of the extrema changes. 
\begin{itemize}
\item The flux value that makes the minimum Minkowski ($n_\textrm{min=Min}$ one might say) is 
\begin{equation}
n^2 = n_{\textrm{Min}}^{\, 2} \equiv \frac{N}{4}.
\end{equation}
For this value, consistent with Eq.~\ref{MinkN},
\begin{equation}
\ \ R^{\, 2}_\textrm{min} = \frac{1}{2} N  \ \ \ \ \textrm{ and } \ \ \ \  V_{\textrm{min}}  = 0 ,
\end{equation}
\begin{equation}
m^{\, 2}_\text{sym} \biggl|_\textrm{min} =  \frac{2}{1+N} \left( \frac{2}{N} \right)^{1+N} \ \ \ \ \textrm{ and } \ \ \ \  m^{\, 2}_\text{asym} \biggl|_\textrm{min}   = 2  \left( \frac{2}{N} \right)^{1+N} .
\end{equation}
At this flux value  the other symmetric extremum is a de Sitter saddle
\begin{equation}
R^{\, 2}_\textrm{max} = 1 + \frac{1}{2} N \ \ \ \ \textrm{ and } \ \ \ \  V_\textrm{max} = \frac{1}{(1 + N/2)^{2 + N} },
\end{equation}
\begin{equation}
m^{\, 2}_\text{sym} \biggl|_\textrm{max}  = -  \frac{2}{1+N} \left( \frac{2}{N+2} \right)^{1+N} \ \ \ \ \textrm{ and } \ \ \ \  m^{\, 2}_\text{asym}  \biggl|_\textrm{max}  = (N-2)  \left( \frac{2}{2+ N} \right)^{2+N} ,
\end{equation}
so that $N=2$ means that $m^{\, 2}_{\textrm{asym}} \bigl|_\textrm{max}  = 0$ for $n=n_{\textrm{Min}}$.
\item When 
\begin{equation}
n^2 = n_{{m^{\, 2}_\text{sym}=0}}^{\, 2} \equiv  \frac{(1 + N)^2}{4(2+N)},
\end{equation}
the $m^{\, 2}_{\textrm{sym}}$ of both the minimum and the symmetric saddle meet at zero, and the two critical points merge and disappear in a fold catastrophe. At this value,
\begin{equation}
R_\textrm{min}^{\, 2} = R_\textrm{max}^{\, 2} = \frac{1 + N}{2} \ \ \ \ \textrm{ and } \ \ \ \ V_\textrm{min} = V_\textrm{max}  = \frac{1}{2+N}\left( \frac{2}{1+N}\right)^{1+N} ,
\end{equation}
\begin{equation}
\ \ \ \ \ \ \ m^{\, 2}_\text{sym} = 0  \ \ \ \ \textrm{ and } \ \ \ \ m^{\, 2}_\text{asym} = \frac{2N}{2+N} \left( \frac{2}{1+N} \right)^{1+N} .
\end{equation}
\item When  
\begin{equation}
n^2 = n^{\, 2}_{m^{\, 2}_\text{asym} = 0} \equiv \frac{2 + 3N}{16}
\end{equation}
 the $m^{\, 2}_\text{asym}$ of the maximum goes through zero. The maximum with $N$ negative eigenvalues becomes a saddle point with just one. 
\end{itemize}
For $N \geq 2$, and with equality only when $N=2$, 
\begin{equation}
n_{m^{\, 2}_\text{asym} = 0}^{\, 2} 
\leq n_{\textrm{Min}}^{\, 2} < n_{m^{\, 2}_\text{sym} = 0}^{\, 2} .
\end{equation} 
That the maximum loses  extra negative modes at the same flux-value that the minimum passes through $V_\textrm{min} = 0$ is a coincidence\footnote{In fact, it's not the entire coincidence.  For $N=2$ there is in some sense a triple coincidence at $n_1=n_2=1/2$.  First, the minimum of the potential has $V=0$.  Second, one of the eigenvalues about the saddle/maximum is zero, $m^2_{\textrm{asym}} = 0$.  Third, the \emph{other} eigenvalue about the saddle/maximum is precisely $m^2_{\textrm{sym}} = -4V/3$, the value where the dominant decompactification instanton switches from being of the Coleman-De Luccia-type to the Hawking-Moss-type, as discussed in \cite{Jensen:1988zx,Hackworth:2004xb,Batra:2006rz,Brown:2011um, Battarra:2013rba}.  This triple coincidence does not persist to larger $N$.} specific to $N=2$.

\subsection{General $\mathbf{N}$ Minkowski}
\label{subsec:GeneralNMinkowski}

The condition to have a Minkowski minimum is given by {Eq.~\ref{MinkN}} as 
\begin{equation}
\sum_{i=1}^N \frac{1}{n_i^{\,2}} = 4.
\end{equation}
The size of the extra dimensions in this minimum is given by Eq.~\ref{MinkN} as 
\begin{equation}
R_i^{\,2} = 2n_i^{\,2}. 
\end{equation}
That these expressions involve no approximations and that for Minkowski minima the value of $R_i$ depends only on $n_i$ makes the $V_4 =0$ case particularly straightforward and instructive. The main work of this section will be deriving the radion masses in the minima. \\

\noindent In Sec.~\ref{sec:minimum} we showed that all Minkowski extrema are necessarily Minkowski minima. Since $V_4=0$ we have
\begin{gather}
V_4 B_i= \frac1{2 n_i^{\,2}}\prod_{k=1}^N\frac1{2 n_k^{\,2}},
\end{gather}
and the Hessian in Eq.~\ref{Hessphi} takes the simple form
\begin{gather}
\partial_{\phi_i} \partial_{\phi_j} V \equiv H^{(\phi)}_{ij}=\left(\prod_{k=1}^N\frac1{2 n_k^{\,2}}\right)\left[\frac1{n_i^{\,2}}\delta_{ij}-\frac A{n_i^{\,2}}-\frac A{n_j^{\,2}} +4A^2\right],
\end{gather}
where $A\equiv {(\sqrt{N+1}-1)}/{(N\sqrt{N+1})}$ as in Eq.~\ref{eq:Adefintion}. The mass squareds of the fluctuations are given by eigenvalues of this $N\times N$ matrix, which are thus the roots of an $N$th order polynomial. The characteristic polynomial is $\det[ H^{(\phi)}_{ij}  - m^2 \,  \mathbf{I} ] = 0$, or (after some algebra, and for brevity writing $\widetilde{m}^2\equiv m^2 \prod_{k=1}^N {2 n^{\, 2}_k}$) 
\begin{eqnarray}
 (\widetilde{m}^2)^N - (\widetilde{m}^2)^{N-1} \frac{N}{N+1}  \sum_i n_i^{-2} + (\widetilde{m}^2)^{N-2} \frac{N-1}{N+1} \sum_{i<j} n_i^{-2}  n_j^{-2}   \hspace{2cm} && \nonumber \\
- (\widetilde{m}^2)^{N-3} \frac{N-2}{N+1} \sum_{i<j<k} n_i^{-2}  n_j^{-2}n_k^{-2}   +  \cdots   +  (-1)^N \frac{1}{1 +N} n_1^{-2} n_2^{-2} \cdots n_{N}^{-2}  &= & 0 . 
\end{eqnarray}
For the special case  $N=2$ this allows us to read off the eigenvalues directly, as
\begin{eqnarray}
m_+^2 \biggl|_{V=0,N=2} & = & \frac{1}{4} \frac{n_1^2+n_2^2 + \sqrt{n_1^4-n_1^2 n_2^2+n_2^4}}{3n_1^{4} n_2^{4}}\\
m_-^2 \biggl|_{V=0,N=2} & = &  \frac{1}{4} \frac{n_1^2+n_2^2 - \sqrt{n_1^4-n_1^2 n_2^2+n_2^4}}{3n_1^{4} n_2^{4}} .
\end{eqnarray}
For any $N$, we can read off the following handsome combinations of eigenvalues 
\begin{eqnarray}
\sum_{i=1}^N  m^{\, 2}_i   &=& \textrm{tr} [ \partial_{\phi_i} \partial_{\phi_j} V]  \ \ \   = \  \frac{4N}{N+1} \left( \prod_{k=1}^N \frac{1}{2 n^{\, 2}_k} \right)  \\
\prod_{i=1}^N  m^{\, 2}_i  &=& \det [ \partial_{\phi_i} \partial_{\phi_j} V] \ = \  \frac{1}{{N+1}}  \left( \prod_{k=1}^N \frac{1}{n^{\, 2}_k} \right) \left( \prod_{k=1}^N \frac{1}{2 n^{\, 2}_k} \right)^{N} \label{eq:detMinkowski} \\ 
\sum_{i=1}^N \frac{1}{  m^{\, 2}_i } & = & \lim_{{m^2} \rightarrow 0} \partial_{m^2} \log [\textrm{characteristic polynomial}] \ = \ \left( 2 \sum_{k=1}^N n_k^2 \right)  \left( \prod_{k=1}^N {\frac{1}{2 n^{\, 2}_k}} \right) . \label{inverseeigenvalues}
\end{eqnarray}

\noindent Let's look at two large-flux limits in which the expressions for the individual eigenvalues simplify.

\subsubsection{Large-flux Minkowski limit 1:  $\mathbf {\{ n_1, n_2, \cdots , n_{N-1} \} \ll n_N }$}

The smallest eigenvalue is 
\begin{equation}
 m_\textrm{lightest}^{\, 2} = \frac{1}{2n_N^2}  \prod_{k=1}^N \frac{1}{2 n^{\, 2}_k} .
\end{equation}
In this limit, the smallest eigenvalue corresponds to changing $R_N$ while leaving all the other $R_i$ unchanged---moving along the `trench' in the effective potential visible in panel B of Fig.~\ref{fig:ToposN2}. 
When there are hierarchies between every flux, $n_1 \ll n_2 \ll \cdots \ll n_N$, the full spectrum is
\begin{equation}
 m_{\textrm{lightest}}^{\, 2}  =  \frac{1}{2 n_N^2} \prod_{k=1}^N \frac{1}{2 n^{\, 2}_k} ; \ \    m_{\textrm{$2^{\textrm{nd}}$-lightest}}^{\, 2}  =  \frac{2}{3n_{N-1}^{\, 2}}  \prod_{k=1}^N \frac{1}{2 n^{\, 2}_k} ;   \  m_{\textrm{heaviest}}^{\, 2} = \frac{N}{(N+1)n_1^2} \prod_{k=1}^N \frac{1}{2 n^{\, 2}_k}.
 \label{eq:MinkFullHierarchy}
\end{equation}

\subsubsection{Large-flux Minkowski limit 2:  $\mathbf{n_1 \ll n_2 = n_3 = \cdots = n_N = n}$ }
\label{sec:minkowskilimit2}
When $n_1$ is only just above $1/2$, all the other $n_i$ must be large for the minimum to be Minkowski. In this subsubsection, we will consider the case where they are not only large but equal. This is a combined hierarchical limit (for $i =1$) and diagonal limit (for $i \neq 1$). 
In this limit there are three distinct eigenvalues. 
The largest eigenvalue corresponds to changing $R_1$,
\begin{equation}
m_{\textrm{heaviest}}^{\, 2} = \frac{4N}{(N+1)} \prod_{k=1}^N \frac{1}{2 n^{\, 2}_k}.
\end{equation}
There are $N-2$ eigenvalues that asymmetrically change the sizes of the $N-1$ large dimensions while keeping the total volume fixed
\begin{equation}
m_\textrm{asymmetric}^{\, 2} = \frac{1}{n^2} \prod_{k=1}^N \frac{1}{2 n^{\, 2}_k}.
\end{equation}
The smallest eigenvalue  corresponds to  symmetrically changing the size of the $N-1$ large dimensions 
\begin{equation}
m_\textrm{symmetric}^{\, 2} =   \frac{1}{N} \frac{1}{n^2} \prod_{k=1}^N \frac{1}{2 n^{\, 2}_k}.
\end{equation}

\subsection{Large-flux de Sitter} 
\label{subsec:generalNdStail}

Since in the large-flux limit all the de Sitter minima lie close to Minkowski minima (as is visible in Fig.~\ref{EquipotentialsN2}), our approach will be to start with the results of Sec.~\ref{subsec:GeneralNMinkowski} and perturb. Here we quote the results, which are derived in the Appendix.

\subsubsection{Large-flux de Sitter limit 1:  $\mathbf {\{ n_1, n_2, \cdots, n_{N-1} \} \ll n_N }$}
\label{sec:largefluxdesitterlimit1}
To leading order in $K n_{i \neq N}^{\, 2}/n_N^2$ and $n_{i \neq N}^{\, 2}/n_N^2$, the sizes of the extra dimensions are 
 \begin{eqnarray}
\hspace{1cm} R_i^2  &=  & {2} n_i^2    \hspace{1cm}  \ \ \ \ \ \ \ \ \ \ \textrm{ for } i \neq N \label{eq:othersizetail} \\
\hspace{1cm} R_N^2 & = &   \frac{{2 n_N^2} ({2-\sqrt{4-3 K}}) }{ K } , \label{eq:sizetail}
\end{eqnarray}
and the value of the potential in the minimum is 
\begin{equation}
V_{\textrm{min}} = \frac{\left(1 - \sqrt{4-3 K}\right) \left(2 + \sqrt{4-3 K}\right)^2}{54 n_N^2} \prod_{i=1}^N \frac{1}{2 n_i^2}. \label{eq:VdStail}  
\end{equation}
In these expressions $K$ is defined to be how much bigger $n_N$ is than it needs to be in order to engender a Minkowski minimum:
\begin{equation}
K \equiv \frac{n^{\, 2}_N}{\bar{n}^{\, 2}_N} \ \textrm{ where } \ \frac{1}{\bar{n}_N^2} \equiv 4 - \sum_{i \neq N} \frac{1}{n_i^2} , \label{Kdefinition}
\end{equation}
so that  $K<1$ is AdS and $1 < K  < {4/3}$ is de Sitter. The value of the lightest radion mass in this minimum is 
\begin{equation}
m_{ \textrm{lightest}}^{\, 2} =  \frac{( 2 + \sqrt{4 - 3K} )^2  \sqrt{4 - 3K} }{9} \frac{1}{2}   \frac{1}{n_N^2}  \prod_{i=1}^N \frac{1}{2 n^{\, 2}_i} .
\end{equation}
The lightest radion mass interpolates from its Minkowski value at $K=1$ down to zero at the fold catastrophe at $K = {4/3}$, where the minimum merges with a saddle and disappears.

\subsubsection{Large-flux de Sitter limit 2: $\mathbf{n_1\ll n_2 = n_3 =\cdots = n_N \equiv n}$}
\label{sec:largefluxdesitterlimit2}
In this regime $R_1 \ll R_2 = R_3 = \cdots = R_N \equiv R$. To leading order in $L n_1^2/n^2$ and $n_1^2/n^2$,
 \begin{eqnarray}
R_1^2 &=& {2n_1^2} , \label{eq:whydoesAlexwantthis} \\
R^2& =& \frac{2n^2(N - \sqrt{N^2 -L N^2 +L}) }{ (N-1)L}, \label{eq:sizetailother} \\
	 V &=& \frac{\left( N + \sqrt{N^2 -L N^2 +L} \right) \left( 1 - \sqrt{N^2 -L N^2 +L} \right) }{2 n^2 (1+N)^2} 	  \prod_{i=1}^N {1 \over R_i^2}.  \label{eq:VdStailother}
 \end{eqnarray} 
In these expressions $L$ is defined to be how much bigger $n$ is than it needs to be in order to beget a Minkowski minimum:
\begin{equation}
L \equiv \frac{n^2}{\bar{n}^{2}} \ \textrm{ where } \ 
	\frac{N-1}{\bar{n}^2} + \frac1{n_1^2}=4, \label{eq:Ldefinition}
\end{equation}
so that $L<1$ is AdS and $1 < L \le {N^2}/{({N^2-1})}$ is de Sitter. For $L=  1$ we recover the Minkowski results of Sec.~\ref{subsec:GeneralNMinkowski}.  The smallest eigenvalue is 
\begin{equation}
m_\textrm{sym}^{\, 2} =   \frac{2 \sqrt{L+N^2-L N^2}}{N R^2} \prod_{i=1}^N {1 \over R_i^2}~,
\end{equation}
which corresponds to symmetrically changing the sizes of the large extra dimensions. There are $N-2$ degenerate radion masses given  by 
\begin{equation}
	m^{\, 2}_{\textrm{asym}} = \frac{2 \left(N-1 + 2\sqrt{N^2 + L - L N^2} \right) }{(N+1) R^2} \prod_{i=1}^N{1 \over R_i^2},
\end{equation}
which correspond to changing the relative sizes of the large extra dimensions while keeping the overall volume fixed. The only eigenvector that changes the volume of the smallest sphere is that associated with the largest eigenvalue
 \begin{equation}
	m_{\textrm{heaviest}}^{\, 2} =  \frac{N}{(1+N) n_1^2} \prod_{i=1}^N {1 \over R_i^2}. 
 \end{equation}\\

The approximations of this subsection cover all the large-flux de Sitter vacua, but also some of the large-flux AdS vacua. For $1<K<{4/3}$ and $1 < L \le {N^2}/(N^2-1)$ the minima are de Sitter, but nothing goes wrong with our approximations if we let $K$ or $L$ become less than 1 (which gives  AdS minima with $n_1$ just above a half---just to the right of the dashed vertical line in Fig.~\ref{EquipotentialsN2}) or let $K$ or $L$ become negative (which gives  AdS minima with $n_1$ just less than a half---just to the left of the dashed vertical line). The approximations of this section thus cover those AdS minima that lie close to the Minkowski surface. However, for AdS minima with  negative $K$ or $L$ that is too large, our expansion breaks down and we need a new approximation. This will be the topic of the next subsection.

\subsection{Large-flux anti-de Sitter} 
\label{subsec:generalNnohumpAdS}

Minima with $n_1 < \frac{1}{2} \ll n_{i \neq 1}$ are all large-flux  AdS minima. To leading order in $1/n_{i \neq 1}$ and in $1/n_{i \neq 1}^{\, 2} \widetilde{V}$, it is derived in the Appendix that 
\begin{eqnarray}
R_{i \neq 1}^{\, 2} &=& \frac{(3 + N)n_i}{2 \sqrt{1 - 4n_1^2 } } \left(1-\sqrt{1-\frac{16 (N+2)}{(N+3)^2} n_1^2 }\, \right)^\frac{1}{2}  \left(\frac{1+N}{3+N}+\sqrt{1-\frac{16 (N+2)}{(N+3)^2} n_1^2 }\, \right)^\frac{1}{2} \label{eq:nohumpeRitail} \\
R_1^2 & = & \frac{N+3}{4} \left(1-\sqrt{1-\frac{16 (N+2)}{(N+3)^2} n_1^2 }\, \right) \label{eq:nohumpeR1tail},  \\ 
V & = & -  \frac{2n_{i \neq 1}^{\, 2}}{ R_{i \neq 1}^4} \prod_{i=1}^N \frac{1}{R_i^2}.  \label{eq:nohumpVtail}  
\end{eqnarray}
Notice that the quantity $n_{i\neq1}^{\,2}/R_{i\neq1}^{\,4}$ is the same for all $i\neq1$.
The approximation that $1/n_{i \neq 1}^{\, 2} \widetilde{V}$ is small is equivalent to the approximation that the curvature contributions, $-R_{i \neq 1}^{-2}$, to the effective potential may be neglected.   This approximation is generally valid in the large-flux limit, and only breaks down when $n_1$ gets too close to a half---when $n_1$ gets too close to a half then Eq.~\ref{eq:nohumpVtail} shows that there is an accidental cancellation between the other terms in the potential and the curvatures cannot be neglected.  
 
 If at the same time as taking $n_{i \neq 1}$ huge, $n_1$ is taken to be tiny the potential scales as 
\begin{equation}
V \sim -  \frac{1}{n_2 n_3 n_4 \cdots n_{N}} \frac{1}{n_1^{N+3}} \label{eq:potentialAdStail}.
\end{equation}
This  expression describes the behavior of a negative equipotential surface in the large-flux limit. 

The $(N-2)$ eigenvectors that correspond to changing the relative sizes of the large spheres while keeping the small one fixed have eigenvalue
\begin{equation}
	 m_{\rm asym}^{\, 2}=  -4  V.
\end{equation}
The other two eigenvectors couple the total-volume mode to fluctuations in the size of the small sphere, and have eigenvalues
\begin{equation}
	m_{\pm}^{\, 2}= \frac{-(3N+4) R_1^2 \widetilde V+N\pm\sqrt{N^2 \left(R_1^2 \widetilde V-1\right)^2+4 R_1^2 \widetilde V}}{(N+1) R_1^2} \prod_{i=1}^N {1 \over R_i^2}. \label{eq:fuam}
\end{equation}
Unlike for the large-flux de Sitter vacua, all $N$ of these eigenvalues have approximately the same scale despite the hierarchy amongst the $n_i$. This feature can be understood in the context of sequential compactification, which we will discuss in the next subsection.

\subsection{Scaling of the large-flux minima and sequential compactification}

The large-flux de Sitter vacua studied in Sec.~\ref{subsec:generalNdStail} and the large-flux AdS vacua studied in Sec.~\ref{subsec:generalNnohumpAdS} have different scalings with $n_i$. The sequential compactification analysis of Sec.~\ref{subsec:sequential} can be used to trace this back to the different scalings of the $N=1$ formulae in Fig.~\ref{fig:TableN1}.

First let's look at the large-flux AdS vacua. In Eqs.~\ref{eq:nohumpeRitail}-\ref{eq:nohumpVtail} we found that the large-flux AdS minima scale with large $n_{i \neq 1}$ as 
\begin{equation}
R_{1}^{\,2} \sim n_{i\neq1}^{\,0} \ ; \  \ R_{i\neq1}^{\,2} \sim n_{i\neq1}   \ ;\  \ \widetilde{V} \sim -n_{i\neq1}^{\,0}. \label{eq:96}
\end{equation}
Since $n_{i \neq 1} \gg n_1$ and $R_{i \neq 1} \gg R_1$ there is a separation of scales that allows us to think about the compactification in two steps. The first step is to compactify on sphere 1, and because $n_1 < 1/2$ this compactification gives an effective $\Lambda_{D-2}$ that is negative. (Indeed, the value of $R_1$ given in Eq.~\ref{eq:nohumpeR1tail} is exactly the value appropriate to compactifying $D$-dimensional spacetime on a single two-sphere, which is why for $D=6$ it reduces to Eq.~\ref{N1Lambdanegmin}.) The second step is to compactify the remaining $N-1$ spheres, a compactification that takes place within the context of a negative  effective $\Lambda_{D-2}$. In Sec.~\ref{subsec:generalNnohumpAdS} we looked at those AdS minima for which this effective cosmological constant is so negative that $|\Lambda_{D-2} n_{i\neq1}^{\, 2}| \gg1$; for these minima the  cosmological-constant term balances against the flux terms, and the curvature terms ($-R^{-2}_{i \neq 1}$) can be neglected. The curvature terms are the only terms that produce backreaction between the various spheres \cite{Brown:2013mwa}, so with no curvature terms the large spheres decouple from one another. Thus, it is like we have an individual $N=1$ compactification for each of the large spheres---the size of each of the large spheres is determined solely by the interplay of its own flux and the large negative effective $\Lambda_{D-2}$. Each compactification has the scaling appropriate to an $N=1$ compactification with $|\Lambda n^2 |\gg 1$, and Eq.~\ref{eq:96} matches the middle column of Fig.~\ref{fig:TableN1}. 

This decoupling also explains why were able to take a more expansive set of approximations for the large-flux AdS vacua than was possible for the large-flux dS vacua. For the large-flux dS minima with $n_1 \ll n_{i \neq 1}$ we were only able to get simple analytic answers when all the $n_{i \neq 1}$ were equal or when there was a large hierarchy between them. For the AdS minima, by contrast, because of the decoupling we needed to assume nothing about the relative sizes of the $n_{i \neq 1}$.

Now let's look at the large-flux de Sitter vacua. In Sec.~\ref{sec:largefluxdesitterlimit1} we looked at those large-flux de Sitter minima with $n_N \gg n_{i \neq N}$. For these minima, the sequential-compactification analysis can again be used---since $R_N$ is so much larger than the other spheres, we can consider first compactifying down to six dimensions and then doing an $N=1$ compactification on the final sphere. In this case, the first $N-1$ compactifications must give rise to an effective $\Lambda_6$ that is positive, because otherwise the final $N=1$ compactification has no chance of producing de Sitter vacua. Recall that in the $N=1$  case with $\Lambda_6 > 0$, small $n$ gives rise to AdS vacua,  intermediate $n$ gives dS vacua and  large $n$ gives no vacua at all: to get a de Sitter vacuum requires that all three terms in the potential---flux, curvature and cosmological constant---are approximately equal. The same is true for the final compactification from six to four dimensions in Sec.~\ref{sec:largefluxdesitterlimit1}: small $K$ gives rise to AdS vacua, intermediate $K$ gives dS vacua, and large $K$ gives no vacua at all. The  vacua with $|K| \ll 1$ have an effective six-dimensional cosmological constant that is so small that it makes a negligible contribution to the potential; these are like $N=1$ compactifications with $\Lambda_6 n_N^{\,2} \ll 1$. Equations~\ref{eq:othersizetail}-\ref{eq:VdStail} give a scaling with $n_N$ in the joint limit $K\ll 1$ and $n_{i \neq N} \ll n_N$ of 
\begin{equation}
R_{i \neq N} \sim n_N^0 \ ; \ R_{N} \sim n_N^2 \ ; \ \widetilde{V}_4 \sim - \frac{1}{n_N^{\, 2}} , \label{eq:97}
\end{equation}
exactly matching the left column of Fig.~\ref{fig:TableN1}. 

\newpage

The scalings of Eq.~\ref{eq:96} are quite different from those of Eq.~\ref{eq:97}. To appreciate this difference, consider what happens when a large-flux de Sitter vacuum loses a few percent of its flux. If  it loses a few percent of the largest flux, $n_N$, then Eqs.~\ref{eq:othersizetail}-\ref{eq:VdStail} tell us that not much happens---$R_1$ is unchanged, $R_N$ shrinks by a few percent, and $V_4$ changes by a few percent. By contrast, if it loses a few percent of the smallest flux, $n_1$, then Eqs.~\ref{eq:nohumpeRitail}-\ref{eq:nohumpVtail}  tell us that the response is dramatic---while $R_1$ shrinks only by a few percent, $R_N$ shrinks dramatically (it now scales as $\sqrt{n_N}$ rather than $n_N$), and the potential is multiplied by a large negative number.

\section{Statistics of Vacua}
\label{sec:statistics}

In this section we count the vacua.  One of the distinguishing features of our model is that there are vacua with arbitrarily many of units of flux around any given cycle. This means that there are an infinite total number of vacua, and that this infinity lies in the large-flux regime of Secs.~\ref{subsec:GeneralNMinkowski}-\ref{subsec:generalNnohumpAdS}. In particular we will see that this gives rise to a pile-up of vacua near $V_4 = 0$, both of de Sitter vacua at $0^+$ and of AdS vacua at $0^-$.

Flux is quantized as in Eq.~\ref{eq:quantization},
\begin{equation}
n_i=\frac{g}{4\pi} \sqrt{\frac{\Lambda_D}{2}}\,\mathbb{Z},
\end{equation}
so the allowed values of the flux occupy grid points in Fig.~\ref{EquipotentialsN2}; the grid spacing  is denser for smaller $g$ and smaller $\Lambda_D$. There will be two steps in computing the distribution of vacua. The first step will be to ignore this quantization: we will use volume in $n_i$-space (in units of the grid-spacing) as a proxy for the number of enclosed grid points. This approximation, discussed in Sec.~\ref{Dirac}, becomes increasingly valid for small values of $g \sqrt{\Lambda_D}$. The second step will be to incorporate the effects of quantization.

\subsection{Statistics of de Sitter minima}
We will see that the volume of $n_i$-space that encloses de Sitter minima of a given $V_4$ is at least
\begin{equation}
\frac{d\hspace{.2mm} \text{Vol}_{\textrm{flux}}}{dV_4}  \sim \left(  V_4 \right)^{-\frac{3(N-1)}{2N}}. \label{eq:dNumdVfordS}
\end{equation}
This has a singularity as $V_4 \rightarrow 0^+$. For $N=2$ the singularity is integrable, which means there's a finite area between the blue and green curves in Fig.~\ref{fig:BPEqui}. For $N=3$ the singularity is logarithmically divergent and for $N \geq 4$ it is power-law divergent. The effect of quantization is generically to replace this  infinity with a double-exponentially large number.

\subsubsection{Flux-space volume}

In this subsubsection we will derive Eq.~\ref{eq:dNumdVfordS}.  We can lower-bound the flux-space volume by considering only those large-flux de Sitter vacua that lie within the remit of Sec.~\ref{sec:largefluxdesitterlimit2}, where $n_1$ is taken to be small and all the other fluxes are taken to be large and approximately equal. 

\newpage

Consider constructing a box in flux space around the point that has $n_2 =n_3 = \dots =n_N \equiv  n$ and has potential $V_4$. How large can we make this box while still requiring that everywhere inside it  has a potential within a few percent of $V_4$? Equation~\ref{eq:Ldefinition} shows that consistent with this requirement we can make $n_{i \neq 1}$ a few percent larger, so that $\Delta n_{i \neq 1} \sim n_{i \neq 1}$. The side of the box corresponding to the smallest flux $n_1$ we can extend less far---Eq.~\ref{eq:Ldefinition} shows that we can only change $n_1$ by an amount $\Delta n_1 \sim (n_1 - \frac{1}{2})  \sim n^{-2} $. In total, then, the flux-space volume of our box is
\begin{equation}
\textrm{Vol}_{\textrm{flux}} \sim \frac{1}{n^2} n^{N-1} \sim n^{N-3}. \label{eq:XXX}
\end{equation}
For $N \geq 3$ most of the volume lives at large $n$, so the largest box with a given value of $V_4$ lives at the largest possible value of $n$. The largest $n$ for a  given value of the potential  is bounded by decompactification, which occurs when $L = N^2/(N^2-1)$, so Eq.~\ref{eq:VdStailother} tells us that we should construct our box at 
\begin{equation}
n_\textrm{max} \sim \left( V_4 \right)^{- \frac{1}{2N}}. \label{eq:YYY}
\end{equation}
We can  insert Eq.~\ref{eq:YYY} into Eq.~\ref{eq:XXX} to get
\begin{equation}
\frac{d \hspace{.2mm} \textrm{Vol}_{\textrm{flux}} }{d \log V_4} \sim n_{\textrm{max}}^{\, N-3}
\ \rightarrow \ 
\frac{d \hspace{.2mm} \textrm{Vol}_{\textrm{flux}} }{dV_4} \sim \left(  V_4 \right)^{-\frac{3(N-1)}{2N}}.
\end{equation}
 This establishes Eq.~\ref{eq:dNumdVfordS} as a lower bound---the flux-space volume must be at least this divergent.  In \cite{Brown:2013fba} we calculated the flux-space volume that arises from a different region of parameter space (specifically the multiply hierarchical limit $n_1 \ll n_2 \ll \cdots \ll n_N$) and found exactly the same divergence. It seems reasonable to believe that Eq.~\ref{eq:dNumdVfordS} is not just a lower bound but gives the exact power-law.

\subsubsection{Enclosed grid points}

For $N \geq 3$, we have shown that the volume of $n_i$-space corresponding to de Sitter minima is infinite. But this infinite volume need not necessarily enclose an infinite number of grid points. The distribution of grid points has a definite structure, a Cartesian grid, and the boundaries of the de Sitter region  align with this grid. Generically the grid of allowed vacua straddles the thin region of large-flux de Sitters, and even though the volume in $n_i$-space is infinite, the number of vacua is finite.

(If we define the function $\# [g]$ to be number of de Sitter minima at a given value of $g$,
then the infinite flux-space volume tells us that the $g$-average of this function $\langle g^N \, \#[g] \rangle_g$ diverges. But it's not that $\# [g]$ is uniformly infinite, rather  the average diverges because of isolated singularities where the axes of grid points happen to coincide with the boundary of the de Sitter region. The quantity we're interested in, however, is not the average of $\# [g]$, but instead the value of $\# [g]$ for our particular value of $g$---and  this will generically be finite.)

The sequential compactification viewpoint gives a simple proof that quantization  does indeed mean that the number of de Sitter minima will typically be finite. For there to have been an infinite number of discrete de Sitter minima, this infinity would have had to have arisen by some subset of the fluxes $(n_1, \dots, n_k )$ all becoming arbitrarily large. This subset can become very large without giving rise to decompactification only if the effective cosmological constant that results from compactifying on the \emph{other} spheres, $\Lambda_{4+2k}$, is tuned to be correspondingly small: in the limit that $(n_1, \dots, n_k )$ become infinitely large, the other spheres must be infinitely tuned to give rise to an effective cosmological constant of exactly $\Lambda_{4 + 2k} = 0^+$. The nonzero level spacing of the $N-k$ small spheres implies that for generic values of $g$ the precise tuning  $\Lambda_{4 + 2k} = 0^+$ will not be achieved and the number of de Sitter minima is finite.
While quantization means there are only a finite number of dS vacua, as $N$ gets large this number becomes huge. In  \cite{Brown:2013fba} we showed the number of de Sitter minima is generically double-exponentially large in $N$.

\subsection{Statistics of AdS minima}

The number of AdS minima is  straightforwardly infinite: the entire strip with $n_1 \leq 1/2$ gives rise to AdS minima for all values of the other $n_i$. Indeed so powerful is this infinity that the flux-volume ${d \hspace{.2mm} \text{Vol}_{\textrm{flux}}}/{dV_4}$ diverges at every negative value of $V_4$, as plotted in Fig.~\ref{fig:BPHisto}.  Flux quantization tempers this divergence so that while there's still an infinite number of AdS minima, the number density only diverges as $V_4 \rightarrow 0^-$.

\subsubsection{Flux-space volume}

Equation \ref{eq:potentialAdStail} gives the equipotential lines for large-flux vacua, $n_1 \ll 1 \ll n_{i \neq 1}$, as
\begin{equation}
n_1^{N+3} n_2 \cdots n_{N-1} n_N = \textrm{constant} \sim - V_4^{-1} .
\end{equation}
The number of large-flux AdS minima with potential between $V_4$ and $2V_4$ is
\begin{equation}
 \int_0 dn_1    \frac{-V_4^{-1}  }{n_1^{N+3}}
  \sim  \infty, \label{eq:divergeAdSvolumedensity}
\end{equation}
so there is a divergence not only in the total flux-volume of AdS minima $\text{Vol}_{\textrm{flux}}$ but also in the differential  $d \hspace{.2mm} \text{Vol}_{\textrm{flux}}/{dV_4}$.

\subsubsection{Enclosed grid points}

All the AdS equipotentials asymptote to $n_i = 0$, so there is some large but finite value of $n_j$ at which the equipotential crosses the innermost plane of nonzero gridpoints. (There are no nonzero grid points for $n_i < {g/4 \pi   \times \sqrt{\Lambda_D/2}}$.) Quantization thus  gives a nonzero lower bound to the integral in Eq.~\ref{eq:divergeAdSvolumedensity} and makes it converge for nonzero $V_4$. The number of AdS vacua with $V_4$ in a given range is thus finite except that there is  an infinite accumulation of vacua as  $V_4 \rightarrow 0^-$.

\section{Discussion}
\label{sec:conclusion}

In this paper, we have investigated flux compactifications on the product of $N$ two-spheres. Each sphere is individually wrapped by a two-form flux, so a vacuum is specified by $N$ integers. We assembled the compactified vacua into a phase diagram, and computed the low-energy effective field theory in each one; there are an infinite number of AdS vacua and a double-exponentially large number of dS vacua. This model has the virtue of being both simple enough to be easily studied and complicated enough to capture interesting new phenomena.

One of these new phenomena is the existence of large-flux de Sitter vacua. There are no large-flux de Sitter vacua in the Bousso-Polchinski model:  because the size and shape of the extra dimensions are held fixed, trying to cram in too many flux lines makes the energy density super-Planckian. But in our model, the extra dimensions are free to react; increasing $n_i$ causes the extra dimensions to expand and, indeed, they expand so much that the energy density actually drops with $n_i$. There are no large-flux de Sitter vacua in the $N=1$ model with $\Lambda > 0$: adding too much flux causes the extra dimensions to decompactify. But in our model, decompactification only happens when there is a large enough flux around every single cycle---if just one of the fluxes is moderate, there can be de Sitter vacua with arbitrarily large flux numbers around the other cycles. Indeed, for $N>2$, the typical de Sitter vacuum has a huge amount of flux around all but one cycle.

The ingredients are simple---gravity, flux, and a positive $\Lambda_D$---but the landscape that has emerged is not. It was once conjectured that a theory with positive $\Lambda_D$ should have $D$-dimensional de Sitter as its state of highest entropy \cite{Banks:2000fe, Bousso:2000nf}. But for our theory this is far from true. Already for the $N=1$ case there are compact de Sitters with entropies larger than that of the parent \cite{Bousso:2002fi}.  For $N \geq 2$, the situation is more dramatic: there are a double-exponentially many compact de Sitter vacua with double-exponentially large entropies.

\subsection*{Acknowledgements}
Thanks to Jose J.~Blanco-Pillado, Raphael Bousso, Frederik Denef, Shamit Kachru, Yasunori Nomura, Eva Silverstein, Paul Steinhardt, Timm Wrase, and Claire Zukowski.

\appendix

\section{Derivation of large-flux results}

In this appendix we provide derivations of the large-flux de Sitter results quoted in Sec.~\ref{subsec:generalNdStail} and of the large-flux anti-de Sitter results quoted in Sec.~\ref{subsec:generalNnohumpAdS}. 

\subsection{Derivation of results in Sec.~\ref{sec:largefluxdesitterlimit1}}

Let us  write Eq.~\ref{eq:Vintermediate} as 
\begin{equation}
\widetilde{V} = \frac{K - 1}{4 n_N^2}  -  \frac{4 n_N^2 \widetilde{V}  + \sqrt{1 - 8 n_N^2 \widetilde{V}} - 1 }{8 n_N^2} - \sum_{i \neq N}  \frac{4 n_i^2 \widetilde{V}  + \sqrt{1 - 8 n_i^2 \widetilde{V}} - 1 }{8 n_i^2} . \label{theequation}
\end{equation}
This equation is exact. Expanding for small $\widetilde{V}$ the last term in the above equation gives
\begin{equation}
- \sum_{i \neq N}  \frac{4 n_i^2 \widetilde{V}  + \sqrt{1 - 8 n_i^2 \widetilde{V}} - 1 }{8 n_i^2}  = \widetilde{V} \sum_{i \neq N} \left( n_i^2 \widetilde{V}  + 4 (n_i^2 \widetilde{V} )^2 + 20 (n_i^2 \widetilde{V} )^3 + \cdots \right) ,
\end{equation}
which will make a negligible contribution to Eq.~\ref{theequation} if $n_{N}$ is much larger than all the other fluxes. Our  approximation therefore is to assume that $n_N \gg n_{i \neq N}$ and drop the last term in Eq.~\ref{theequation}, which leaves 
\begin{equation}
\widetilde{V} = \frac{\left(1 - \sqrt{4-3 K}\right) \left(2 + \sqrt{4-3 K}\right)}{18 n_N^2} + O\left( \frac{n_{i \neq N}^{\, 2}}{n_N^4} \right) . \label{eq:tildeVinthetail} 
\end{equation}
To leading order the only $R_i$ that changes from its Minkowski value is $R_N$; Eqs.~\ref{NRminimum}~\&~\ref{eq:tildeVinthetail} together give  Eqs.~\ref{eq:sizetail} \& \ref{eq:VdStail}. \\

\noindent Substituting the leading-order approximation for $\widetilde{V}$, given by Eq.~\ref{eq:tildeVinthetail}, into Eq.~\ref{eq:doublederivativeVtilde} gives 
\begin{eqnarray}
B_{i \neq N}  &=& \frac{1}{\widetilde{V}} \frac{1}{2n_{i \neq N}^{\, 2}}  \\
B_N &= & \frac{1}{\widetilde{V}} \frac{( 2 + \sqrt{4 - 3K} )(1 + 2 \sqrt{4 - 3K}) }{18n_N^{\, 2}} .
\end{eqnarray}
Then to leading order the trace Eq.~\ref{tracephi} is
\begin{equation}
\textrm{trace  } = \frac{2NV}{N+1}\left(-1+\sum_{k=1}^N B_k\right) =  \frac{4N}{N+1}\frac{V}{\widetilde{V}}  =  \frac{4N}{1+N}  \frac{2+\sqrt{4-3 K}}{3}  \prod_i \frac{1}{2 n_i^{\, 2}} .
\end{equation}
To leading order the determinant Eq.~\ref{Hphidet} is 
\begin{eqnarray}
\det 
& = & \frac{1}{1+N} \left( \frac{  ( 2 + \sqrt{4 - 3K} )  \sqrt{4 - 3K} }{3} \prod_i \frac{1}{n_i^2}  \right) \left( \frac{2+\sqrt{4-3 K}}{3}  \prod_i \frac{1}{2 n_i^2} \right)^N  .
\end{eqnarray}
In the $K \rightarrow 1$ limit, this expression reduces to the Minkowski expression, Eq.~\ref{eq:detMinkowski}, as required. 
Since  $R_N \gg R_{i \neq N}$, changing $K$ does not backreact on the other spheres. The non-smallest eigenvalues shift from their Minkowski values as
\begin{equation}
 m_{i \neq N}^{\, 2}(K)  =  \frac{2 + \sqrt{4 - 3K}}{3}   \times \left(  m_{i \neq N}^{\, 2} \biggl|_{K=1} \right) ,
\end{equation}
so that the dimensionless combination $R_N m_{i \neq N}$ is independent of $K$: the non-smallest eigenvalues only care about the value of $K$ through its impact on the four-dimensional Planck mass. The smallest eigenvalue shifts as 
\begin{eqnarray}
m_{ N}^{\, 2}(K)  & = &  \frac{  ( 2 + \sqrt{4 - 3K} )^2  \sqrt{4 - 3K} }{9}   \times \left(   m_{ N}^{\, 2} \biggl|_{K=1} \right) .
 \end{eqnarray}

\subsection{Derivation of results in Sec.~6.4.2}

Let us  write Eq.~\ref{eq:Vintermediate} as 
\begin{equation}
\widetilde{V} = (N-1)\frac{L - 1}{4 n^2}  - (N-1) \frac{4 n^2 \widetilde{V}  + \sqrt{1 - 8 n^2 \widetilde{V}} - 1 }{8 n^2} -  \frac{4 n_1^2 \widetilde{V}  + \sqrt{1 - 8 n_1^2 \widetilde{V}} - 1 }{8 n_1^2} . \label{theequation2}
\end{equation}
This equation is exact. Expanding for small $\widetilde{V}$ the last term in the above equation gives
\begin{equation}
-   \frac{4 n_1^2 \widetilde{V}  + \sqrt{1 - 8 n_1^2 \widetilde{V}} - 1 }{8 n_1^2}  =   \widetilde{V} \left( n_1^2 \widetilde{V}  + 4 (n_1^2 \widetilde{V} )^2 + 20 (n_1^2 \widetilde{V} )^3 + \cdots  \right) 
\end{equation}
which will make a negligible contribution to Eq.~\ref{theequation2} if $n \gg n_1$. Our  approximation therefore is to assume that $n \gg n_{1}$ and drop the last term in Eq.~\ref{theequation2}, which leaves 
\begin{equation}
\widetilde{V} = \frac{\left( N + \sqrt{N^2 -L N^2 +L} \right) \left( 1 - \sqrt{N^2 -L N^2 +L} \right) }{2 n^2 (1+N)^2} + O\left( \frac{n_{1}^{\, 2}}{n^4} \right) . \label{eq:tildeVintheothertail} 
\end{equation}
To leading order  $R_1$ does not change from its Minkowski value; Eqs.~\ref{NRminimum}~\&~\ref{eq:tildeVintheothertail} together give  Eqs.~\ref{eq:sizetailother} \& \ref{eq:VdStailother}. \\

 \noindent Substituting the leading-order approximation for $\widetilde{V}$, given by Eq.~\ref{eq:tildeVintheothertail}, into Eq.~\ref{eq:doublederivativeVtilde} gives 
\begin{eqnarray}
B_1  &=& \frac{1}{\widetilde{V}} \frac{1}{2n_1^2}  \\
B_2 = B_3 = \dots =  B_N \, \equiv B \ &= & \frac{1}{\widetilde{V}}\frac{N-1+ 2 \sqrt{N^2 + L - N^2 L}}{N+1} \frac{1}{R_{i \neq1}^{\, 2}}  .
\end{eqnarray}
Plugging these expressions into  Eq.~\ref{Hessphi}, the Hessian matrix becomes
 \begin{equation}
        H^{(\phi)}_{ij}=  2V\left[B_i \delta_{ij}- A(B_i + B_j)-\frac1{N+1} + A^2 \left(B_1 + (N-1)B\right)\right],
\end{equation}
and the mass squareds of the radion fluctuations are given by the eigenvalues of this matrix.  As in the Minkowski case (Sec.~\ref{sec:minkowskilimit2}), the lightest eigenvalue corresponds to symmetric fluctuations in which all of the large spheres grow or shrink in unison, and the small sphere is stationary; the eigenvalue corresponding to this eigenvector is given by
\begin{gather}
m_{\textrm{sym}}^{\, 2} =  \frac{2 \sqrt{L+N^2-L N^2}}{N} \frac1{R_1^{\,2} \,R^{2N}}.
\end{gather}
There are also $N-2$ eigenvalues that asymmetrically change the sizes of the $N-1$ large dimensions which keeping the total volume fixed ($\Delta \phi_i- \Delta \phi_j$, for instance) given by
 \begin{equation}
        m^{\, 2}_{\textrm{asym}} = 2V B_{i \neq 1}=\frac{2\left(N-1+ 2\sqrt{N^2 + L - L N^2} \right)}{(N+1)}  \frac1{R_1^{\,2} \,R^{2N}}.
 \end{equation}
 Finally, the heaviest eigenvector, orthogonal to the rest, has eigenvalue
 \begin{gather}
 m_{\textrm{heaviest}}^{\, 2} =  \frac{N}{(1+N) n_1^{\,2}} \frac1{R_1^{\,2} \,R^{2(N-1)}}.
  \end{gather}

\subsection{Derivation of results in Sec.~\ref{subsec:generalNnohumpAdS}}

Let us write Eq.~\ref{eq:Vintermediate} as
\begin{eqnarray}
 \widetilde{V} & = & 1 - \frac{1 + 4 n_1^2 \widetilde{V}  + \sqrt{1 - 8 n_1^2 \widetilde{V}} }{8 n_1^2} - \frac{(N-1) \widetilde{V}}{2} - \sum_{i \neq 1}^N \frac{1  + \sqrt{1 - 8 n_i^2 \widetilde{V}} }{8 n_i^2} .    \label{eq:VintermediateForAdS}
\end{eqnarray}
This equation is exact. Expanding for large $n_i^2 \widetilde{V}$, the last term in the above equation gives
\begin{equation}
- \sum_{i \neq 1}^N \frac{1   + \sqrt{1 - 8 n_i^2 \widetilde{V}} }{8 n_i^2}  = - \frac{\widetilde{V}}{2} \sum_{i \neq 1}^N \left( \frac{1}{\sqrt{-2 n_i^2 \widetilde{V}} }+ \frac{1}{4 n_i^2 \widetilde{V}}  + \cdots \right) .
\end{equation}
For $|n_{i \neq 1}^{\, 2} \widetilde{V}| \gg 1$ this makes a negligible contribution to Eq.~\ref{eq:VintermediateForAdS}, and our  approximation is to drop this term, which leaves 
\begin{equation}
\widetilde{V} = -  \frac{ 3 + N - 8(N+2) n_1^2 +  \sqrt{(3+N)^2 - 16(2+N)n_1^2} }{4 (2+N)^2 n_1^2}  . \label{eq:tildeVintheAdStail} 
\end{equation}
This expression is accurate so long as $n_{i \neq 1} \gg 1$ (so that we are in the large-flux regime), and $|n_{i \neq 1}^{\, 2} \widetilde{V}| \gg 1$ (so that there is not a leading order cancellation in the contributions to $\widetilde{V}$---there is a such a cancellation near $n_1 = 1/2$, where we instead need to use the approximations of Sec.~\ref{subsec:generalNdStail}). This regime is equivalent to being able to neglect the $R_{i \neq 1}^{-2}$ curvature contributions to $\widetilde{V}$. The large $\widetilde{V}$ limit of Eq.~\ref{NRminimum} allows us to approximate
\begin{equation}
R_i^{\,2}  = \frac{\sqrt{ 2} n_i}{\sqrt{- \widetilde{V}}} \left( 1+ O\Bigl( \frac{ \sqrt{-\widetilde{V}}}{n_i} \Bigl)\right) , \label{eq:AdSsizesofotherguys} 
\end{equation}
which together with Eqs.~\ref{NRminimum} \& \ref{eq:tildeVintheAdStail} gives Eqs.~\ref{eq:nohumpeRitail}-\ref{eq:nohumpVtail}. 
In this limit, to leading order,
\begin{eqnarray}
\widetilde V B_1 &=& -2\widetilde V +  {1 \over R_1^2},\\
\widetilde V B_{i\neq 1}&=&-2 \widetilde V.
\end{eqnarray}
Plugging these expressions into  Eq.~\ref{Hessphi}, the Hessian matrix becomes
\begin{equation}
H^{(\phi)}_{ij}= \frac{2}{\prod_{k=1}^N R_k^{\,2}}\left[\frac{\widetilde V}{N+1}+\frac{A^2}{R_1^{\,2}}-2\widetilde V\delta_{ij} +\frac1{R_1^{\,2}}\delta_{i1}\delta_{1j} - \frac{A}{R_1^{\,2}}(\delta_{ij}+\delta_{1j})\right].
\end{equation}
This matrix has an $(N-2)$-fold degenerate mass eigenvalue
\begin{equation}
        m_{\rm asym}^{\, 2}=  -4V,
\end{equation}
with corresponding eigenvectors that fluctuate the large spheres in a volume-preserving way.
The other two eigenvalues are given by Eq.~\ref{eq:fuam}; the corresponding eigenvectors are
\begin{equation}
        V_\pm= \left( \begin{array}{c}
        w_\pm \\
        1 \\
        \vdots\\
        1
        \end{array}
        \right)~,
\end{equation}
where
\begin{eqnarray}
        w_\pm&\equiv& \big[2 N^3+4 N^2+\sqrt{N+1} \left(4-N^2\right)-2 N-2 R_1^2 N^2-4\big]^{-1} \times \cr \cr
         &&\left\{-N^4+(N-2) N^2 R_1^2 +5 N^2+ (2-2N)(2+N)\sqrt{N+1} -4\right. \cr\cr
         && \left. \pm N^2 \sqrt{N^2 R_1^4-2 (N+2) \left(N^2-1\right) R_1^2+(N+2)^2 \left(N^2-1\right)} \right\}~.
\end{eqnarray}

\bibliographystyle{utphys}
\bibliography{mybib.bib}

\end{document}